\documentclass[%
reprint,
nofootinbib,
 amsmath,amssymb,
 aps,
]{revtex4-2}

\usepackage{graphicx}
\usepackage{dcolumn}
\usepackage{bm}
\usepackage{booktabs} 
\usepackage{hyperref}
\usepackage{xcolor}
\definecolor{medium-blue}{rgb}{0,0,1}
\hypersetup{colorlinks, urlcolor={medium-blue}}
\usepackage[font=small,labelfont=bf,
   justification=raggedright,
   format=plain]{caption}

\usepackage{subcaption}
\usepackage{pifont}
\usepackage{float}
\newcommand{\cmark}{\ding{51}}%
\newcommand{\xmark}{\ding{55}}%

\begin{document}

\preprint{APS/123-QED}

\title{cDVGAN: One Flexible Model for Multi-class Gravitational Wave Signal and Glitch Generation}

\author{Tom Dooney$^{1}$}
\author{R. Lyana Curier$^{1}$}
\author{Daniel Stanley Tan$^{1}$}
\author{Melissa Lopez$^{2,3}$}
\author{Chris Van Den Broeck$^{2,3}$}
\author{Stefano Bromuri$^{1}$}

\address{$^1$Faculty of Science, Open Universiteit, Valkenburgerweg 177, 6419 AT Heerlen, The Netherlands}
\address{$^2$Institute for Gravitational and Subatomic Physics (GRASP), Utrecht University, Princetonplein 1, 3584 CC, Utrecht, The Netherlands}
\address{$^3$Nikhef, Science Park 105,
1098 XG, Amsterdam, The Netherlands.}

\begin{abstract}

Simulating realistic time-domain observations of gravitational waves (GWs) and other events of interest in GW detectors, such as transient noise bursts called glitches, can help in advancing GW data analysis. 
Simulated data can be used in downstream data analysis tasks by augmenting datasets for signal searches, balancing data sets for machine learning applications, validating detection schemes and constructing mock data challenges.  
In this work, we present Conditional Derivative GAN (cDVGAN), a novel conditional model in the Generative Adversarial Network framework for simulating multiple classes of time-domain observations that represent gravitational waves (GWs) and detector glitches. cDVGAN can also generate generalized hybrid samples that span the variation between classes through class interpolation in the conditioned class vector.
cDVGAN introduces an additional player into the typical 2-player adversarial game of GANs, where an auxiliary discriminator analyzes the first-order derivative time-series.
Our results show that this provides synthetic data that better captures the features of the original data.
cDVGAN conditions on three classes in the time-domain, two denoised from LIGO \textit{blip} and \textit{tomte} glitch events from its 3rd observing run (O3), and the third representing binary black hole (BBH) mergers. 
Our proposed cDVGAN outperforms 4 different baseline GAN models in replicating the features of the three classes.
Specifically, our experiments show that training convolutional neural networks (CNNs) with our cDVGAN-generated data improves the detection of samples embedded in detector noise beyond the synthetic data from other state-of-the-art GAN models. Our best synthetic dataset yields as much as a 4.2\% increase in \textit{area-under-the-curve} (AUC) performance, maintaining the same CNN architecture, compared to synthetic datasets from baseline GANs.
Moreover, training the CNN with class-interpolated hybrid samples from our cDVGAN outperforms CNNs trained only on the standard classes, when identifying real samples embedded in LIGO detector background between signal-to-noise ratios ranging from 1 to 16 (4\% AUC improvement for cDVGAN).
We also illustrate an application of cDVGAN in a data augmentation example, showing that it is competitive with a traditional augmentation approach.
Lastly, we test cDVGAN's BBH signals in a fitting-factor study, showing that the synthetic signals are generally consistent with the semi-analytical model used to generate the training signals and the corresponding parameter space.

\end{abstract}

\maketitle

\section{\label{sec:Introduction}Introduction}

The first detection of a Gravitational Wave (GW) from a binary black hole (BBH) merger in 2015 ushered in a new era of astronomy and cosmology \cite{first_GW}. Since then, over three observing runs (O1, O2, O3), advanced LIGO \cite{LIGO_paper} and Virgo \cite{VIRGOpaper} detectors have made confident detections of 90 compact binary coalescence (CBC) events, as reported in the Gravitational Wave Transient Catalogues GWTC-1, GWTC-2, GWTC-3 \cite{GWTC1, GWTC2, GWTC3_true}.
With the introduction of KAGRA \cite{KagraPaper}, Japan's underground detector, towards the end of O3 and the O4 run currently underway, hundreds of more detections are expected from the enhanced sensitivity of GW detectors \cite{Detchar_nextgen_glitch}.

Ongoing upgrades to advanced detector systems will give rise to new challenges in Gravitational Wave (GW) data analysis, particularly with the introduction of next-generation GW detectors, such as the Einstein Telescope (ET) \cite{ET_paper} and Cosmic Explorer \cite{Cosmic_expl}.
GW detection rates from all sources are expected to significantly increase i.e. it is estimated the ET will detect on the order of $8 \times 10^4 \, \text{y}^{-1}$ BBH mergers \cite{Iacovelli_2022} and $7 \times 10^4 \, \text{y}^{-1}$ Binary Neutron Star (BNS) mergers \cite{Kalogera_BNS}. 
This could lead to over 400 compact binary coalescence (CBC) events daily. The enhanced detectors are also expected to detect new astrophysical sources of GWs \cite{CCSNa, Neutron_star}.

Aside from genuine astrophysical signals, the heightened sensitivity of GW detectors is expected to exacerbate issues relating to `glitches' \cite{LIGO_second_third_detchar, Virgo_detchar_O3, Kagra_detchara, Detchar_nextgen_glitch}. 
Glitches are non-Gaussian transient noise artefacts that can resemble astrophysical signals, hindering GW data analysis and increasing false positives ~\cite{abbott2018effects, KAGRA:2022dwb,Steltner:2023cfk, KAGRA:2021kbb,Steltner:2021qjy, pankow2018mitigation,davis2019improving,driggers2019improving, blackburn2008lsc, abbott2016characterization}.
Glitches are unmodelled noise events, stemming from environmental or instrumental factors, with some sources remaining unidentified \cite{detchar_transient}.
Unlike modelled CBC events, which are detected through matched-filtering \cite{Owen_1999_match, Usman_2016_match, CANNON2021100680_GstLAL}, glitches are detected by model-free detection algorithms that scrutinize excess power in the time-frequency (spectrogram) representation to distinguish them from the detector background \cite{Omicron}.

To meet the challenges of advanced detector systems, machine learning algorithms have become increasingly popular in the GW physics community \cite{ML_GW_1, ML_GW_2, ML_GW_3, ML_GW_cuoco}.
In the case of glitches, studies have largely focused on spectrogram representations for identification due to their unmodelled nature. 
For example, \textit{Gravity Spy} \cite{Zevin_2017, GspyO4} have made significant strides in characterising spectrogram representations of glitches by combining machine learning and citizen science.
Multiple studies have employed machine learning extensively to improve classification accuracy on \textit{Gravity Spy} spectrograms \cite{gspy_ml_1, gspy_ml_2, gspy_ml_3}. 
Others have leveraged the Generative Advsersarial Network (GAN) \cite{GANpaper} framework for generating \textit{Gravity Spy} spectrograms to augment glitch datasets for further improvement \cite{Jade_Powell_paper, GAN_spec_best}.
However, relying on computationally expensive spectrogram transformations to identify and simulate unmodelled events like glitches may not always be feasible as GW detector technology improves.

In this study, we develop a generative modelling framework for diverse classes of time-domain observations in GW detectors. 
Generating time-domain representations of GWs and glitches offers various advantages, such as their flexibility for experimental purposes and low dimensionality requiring less computational expense.
Simulated data can be used in downstream applications such as data augmentation and class balancing for machine learning applications, validating detection schemes via software injections \cite{Abadie_2010_injection_1, Abadie_2010_injection_2, Biwer_2017_inject} and constructing mock data challenges.
Furthermore, where unmodelled transients like glitches can be isolated from the background, transforming time-series to spectrograms is straightforward, while the reverse is challenging due to background noise captured in spectrograms.

An approach to simulating time-domain glitch events is implemented using the \textit{gengli} glitch generator \cite{GENGLI}.
The authors implement a Wasserstein GAN (WGAN) \cite{wGAN_paper} on blip glitches extracted from detector backgrounds using \textit{BayesWave} \cite{BayesWave}.
They show that it is possible to isolate blips from their surroundings and learn their underlying distribution in the time-domain with WGANs. 
Another study \cite{McGinn_2021} implements a conditional GAN (cGAN) called McGANn to simulate 5 waveform classes analogous to GW bursts.  
Aside from generating five distinct classes, their GAN can generate class-interpolated, hybrid samples.

In this work, we propose a novel conditional Derivative GAN (cDVGAN) that simulates different classes of time-domain observations of LIGO glitch classes and/or astrophysical waveforms.
We condition cDVGAN on three classes; two \textit{Gravity Spy} glitch classes called \textit{blip} and \textit{tomte}, and a third represented by BBH signals.
Aside from learning realistic distributions of diverse classes in one flexible model, cDVGAN can generate hybrid samples that traverse the variation between the learned classes by manipulating the user-controlled class vector.
After the training phase, the models can generate hundreds of thousands of in-class or hybrid samples in a matter of seconds, per the user's preference.
cDVGAN is intended for use in a next-generation glitch generator, once investigations into \textit{BayesWave} or other denoising algorithms can provide reliable time-domain representations of other glitch types.

To show the utility of our generated (synthetic) data, we implement an experiment that uses it for training a convolutional neural network (CNN) detection algorithm to identify real data from the GAN training data distribution in additive LIGO detector noise.
Our results indicate that cDVGAN can better capture the features of the data than other state-of-the-art GAN models by incorporating adversarial feedback on the first-order derivatives using an auxiliary discriminator. 
Furthermore, we show that GAN-generated hybrid samples can be useful for training detection algorithms to identify real data in noise beyond the standard GW and glitch classes.

Since cDVGAN is primarily intended for use as a glitch generator, the astrophysical nature of the BBH signals is considered arbitrary for most of the experiments.
However, we also confirm our cDVGAN data against a semi-analytical model used in GW searches, exploring cDVGAN's application as a BBH signal generator. To this end, we implement a fitting-factor study to evaluate the faithfulness of our generated data against the templates of a template bank. 
The results show that cDVGAN's synthetic signals are generally consistent with signals from the waveform routine used to generate the original training signals.

This paper is structured as follows; in section \ref{sec:Generative_Adversarial_Networks}, we discuss concepts relating to GANs. 
In section \ref{sec:Methods}, we present the cDVGAN architecture and training schemes, the datasets and preprocessing, and experimental details. 
In section \ref{sec:Results}, we present the experimental results involving a simple search for real GWs and glitches in additive detector background using synthetic data from 5 different GAN models. 
Finally, section \ref{sec:Conclusion} discusses the conclusions of this research.

\section{\label{sec:Generative_Adversarial_Networks}Generative Adversarial Networks}

\begin{figure*}
\centering
\captionsetup{width = 0.85 \textwidth}
\includegraphics[width = 0.9\textwidth]{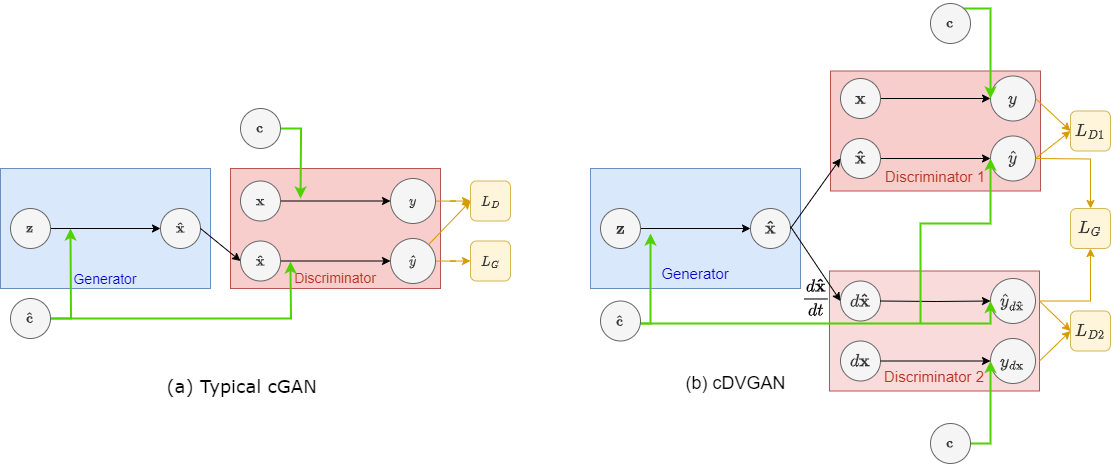}
\caption{Diagrams of a typical cGAN architecture (\textit{left}), comprising one discriminator, and cDVGAN (\textit{right}), comprising two discriminators. Class vectors $c$ (real) and $\hat{c}$ (fake), are fed to all model components in both cases. An intermediate derivative calculation is observed in the cDVGAN plot, where the derivative of the synthetic sample is calculated. cDVGAN2 includes yet another discriminator applied to second-order derivatives. In cDVGAN and cDVGAN2, the total generator loss is calculated as a linear combination of the discriminator losses applied to synthetic samples.}
\label{fig:dual_discriminator_gan}
\end{figure*}

\subsection{Wasserstein GANs}

GANs are a class of machine learning algorithms that generate realistic synthetic data. 
They consist of two neural networks: a discriminator (also known as a critic) that distinguishes between real and synthetic data; and a generator that generates synthetic data that can fool the discriminator. 
GAN approaches suffer from stability issues such as vanishing gradients stimulating numerous studies centred around methods to stabilize the training process.
Wasserstein GAN \cite{wGAN_paper} is a particular variant that addresses these issues.
It uses the Wasserstein-1 distance ($W1$) as the loss function to measure the similarity between the real and synthetic distributions.
$W1$ is fully differentiable and increases monotonically while never saturating, removing the issue of vanishing gradients. 
Under this paradigm, the optimization problem can be formulated as 

\begin{equation}\label{eq_wGAN_opt}
    \theta_{opt} = arg \min_{\theta}\max_{\phi: ||D(x, \phi)||_{l} \leq 1}L(\phi, \theta)
\end{equation}
where the maximum is taken over all 1-Lipschitz functions $D$ and with $L$ defined as
\begin{equation}\label{Discriminator_loss}
    L(\phi, \theta) = \mathbb{E}_{x \sim P_x}[D(x, \phi)] - \mathbb{E}_{\hat{x} \sim P_{\hat{x}}}[D(\hat{x}, \phi)]
\end{equation}
where $\hat{x} = G(z, \theta)$ and $z$ is a batch of the generator's latent vector. 
$D$ and $G$ refer to the discriminator and the generator with parameters $\phi$ and $\theta$, respectively. $\mathbb{E}_{x \sim P_x}$ averages over a batch of real samples $x$ from the real distribution $P_x$, while $\mathbb{E}_{\hat{x} \sim P_{\hat{x}}}$ averages over a batch of generated samples $\hat{x}$ from the synthetic distribution $P_{\hat{x}}$. 
Equation \ref{eq_wGAN_opt} requires a constraint of 1-Lipschitz continuity on $D$ \cite{wGAN_paper}. 
This can be accomplished by adding a regularization penalty called the gradient penalty ($GP$) to the discriminator loss \cite{wGAN_GP_paper, improved_GAN_training, karras2018progressive}. 
The discriminator loss then becomes 
\begin{equation} \label{eq_wGAN_gp}
    L_D = - L(\phi, \theta) + \lambda GP(\phi)
\end{equation}
with 
\begin{equation}\label{eq_GP}
    GP(\phi) = \mathbb{E}_{\hat{x} \sim P_{\hat{x}}} \left[(||\nabla_{x} D(\hat{x}, \phi)||_2 - 1 )^2\right]
\end{equation}

and where $\lambda$ represents the regularization hyperparameter, $||.||_2$
represents the $L^2$-norm and $\hat{x}$ is a randomly sampled point between the real and synthetic data. 
When updating the generator, errors are propagated through the entire network, from $D$ to $G$. Naturally, for the generator updates are made only on generated samples from $G$. The generator loss is written as
\begin{equation}\label{eq:generator_loss}
    L_G(\phi, \theta) = -\mathbb{E}_{\hat{x} \sim P_{\hat{x}}}[D(\hat{x}, \phi)]
\end{equation}

\begin{figure}[H]
\centering
\captionsetup{width = 0.45 \textwidth}
  \includegraphics[width=0.4\textwidth]{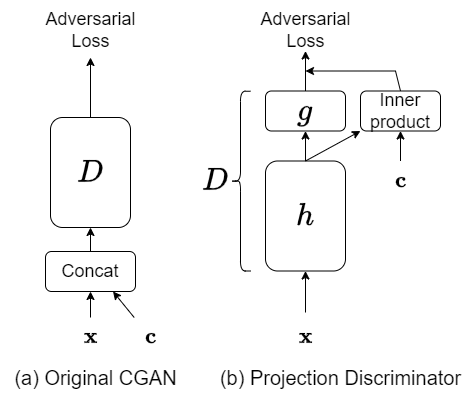}
\caption{A comparison of the discriminators from the original cGAN paper (used for McGANn and McDVGANn) and the projection discriminator (used for cWGAN, cDVGAN, cDVGAN2).}
\label{fig:conditioning_methods}
\end{figure}

\subsection{Conditional GANs \label{sec:conditional_gan}}
Conditional GANs (cGANs) \cite{CGAN_paper} allow finer control over the generated data by providing extra information to both the generator and discriminator. 
For example, we can specify the class of the generated data by providing class label, $c$. 
The training data and class labels are taken from a joint distribution $P_{data}(x, c)$. 
When generating synthetic samples, the class vector $\hat{c}$ and the GAN latent vector distribution $P_{z}$, are sampled independently.

In the original cGAN paper \cite{CGAN_paper}, conditional information was provided to the network by naively concatenating the class information to the first layers of the GAN components. 
In the case of time-series, this corresponds to concatenating it to the $z$ vector before passing it to the generator, and the input sample before passing it to the discriminator. 

Miyato and Koyama \cite{cgan_projection} developed a more effective conditioning method using a projection in the discriminator between the conditional features and the features extracted from the input before the discriminator output. 
The projection output measures the similarity between the condition and the discriminator's feature vector and is added to the discriminator output. 
The output of the discriminator is modified from

\begin{equation}
    D(x, c) = g(h(x,c))
\end{equation}
 to
\begin{equation}\label{eq:projection_condition}
    D(x,c) = c^TVh(x) + g(h(x))
\end{equation}
where $V$ is an embedding matrix, $h$ is the discriminator's feature vector and $g$ is the discriminator output layer (see Figure \ref{fig:conditioning_methods}). This approach has been shown to improve the quality of class conditional generation using GANs.
Note that the generator is still conditioned by concatenating the class-embedded vector to the latent input, similar to the original cGAN paper.

\section{\label{sec:Methods}Methods}

\subsection{\label{sec:cDVGAN_method}Conditional Derivative GAN (cDVGAN)}

The adversarial training process for GANs is known to be volatile. Models can suffer mode collapse, where only a few realistic samples are learned, while they can also fail to converge at all. These issues often occur when one model component (generator/discriminator) begins to dominate the other during the training phase. These problems are exacerbated when conditioning on multi-modal distributions of GW and glitch time-series.  In this section, we introduce two new cGAN designs, cDVGAN and cDVGAN2, that can help overcome these limitations. 

In cDVGAN, two discriminators are applied to two different representations of the data instead of the usual single discriminator. 
The first discriminator is applied to the original samples as in a conventional GAN, while the second discriminator is applied to the corresponding derivative samples.
A high-level diagram of cDVGAN can be seen in Figure \ref{fig:dual_discriminator_gan}.
We have shown in our previous work that the derivative discriminator leads to increased training stability of the model components and minimizes high-frequency artefacts in the GAN output, generating smoother and more faithful data \cite{dooney2022dvgan}.
Under this scheme, the generator loss is calculated as a linear combination of the two discriminator outputs. Equation \ref{eq:generator_loss} can be rewritten as 
\begin{multline}\label{eq:combined_generator}
     L_G(\phi_1, \phi_2, \theta) = -\eta_1\mathbb{E}_{\hat{x}_1 \sim P_{\hat{x}_1}}[D_1(\hat{x}_1, \phi_1, \hat{c})]\\-\eta_2\mathbb{E}_{\hat{x}_2 \sim P_{\hat{x}_2}}[D_2(\hat{x}_2, \phi_2, \hat{c})]
\end{multline}

where $D_1$ and $D_2$, represent the first and second discriminator respectively. 
Here $\hat{x}_1$ represents synthetic samples while $\hat{x}_2 = d\hat{x}_1/dt$ represents the corresponding derivatives and $\hat{c}$ represents the class vector for the synthetic samples.
$P_{\hat{x}_1}$ and $P_{\hat{x}_2}$ are the distributions of the two representations of generated data, and $\eta_1$ and $\eta_2$ are hyperparameters that control the relative strength of the discriminator losses.
The cDVGAN method is not restricted to two discriminators.
We extend cDVGAN to cDVGAN2, which includes a third discriminator that is applied to second-order derivatives.
The models are identical except for the addition of the second-order derivative discriminator in cDVGAN2, while the second-order derivative discriminator is identical to the first-order discriminator except for the input size.
Additional representations of the data (eg. time-frequency representations) can also be provided to additional discriminators, depending on the problem.
For k discriminators applied to k representations of the data, the generator loss is written generally as 

\begin{equation}\label{eq:multi_wgan}
    L_G(\phi_1,...,\phi_k, \theta) = -\sum_{i=1}^k \eta_i\mathbb{E}_{\hat{x}_i \sim P_{\hat{x}_i}}[D_i(\hat{x}_i, \phi_i, \hat{c})]
\end{equation}

Our cDVGAN and cDVGAN2 models are conditioned via projection, as described in section \ref{sec:conditional_gan}. 
Since hyperparameter optimization is not the focus of this research, the hyperparameters $\eta_1 = \eta_2 = 0.5$ in cDVGAN and $\eta_1 = \eta_2 =\eta_3 = 0.33$ in cDVGAN2, meaning all discriminators contribute equally to the respective generator losses.
The discriminators are updated 5 times for each generator update.
The full model architecture can be viewed in the Appendix (\ref{sec:cDVGAN_Architecture})\footnote{Python code can be found at \href{https://git.ligo.org/tom.dooney/cdvgan_paper}{https://git.ligo.org/tom.dooney/cdvgan\_paper}.}.

\subsection{\label{sec:Data_preprocessing}Training Data and Preprocessing}

\begin{figure}
\centering
\captionsetup{width = 0.5 \textwidth}
\includegraphics[width = 0.5\textwidth]{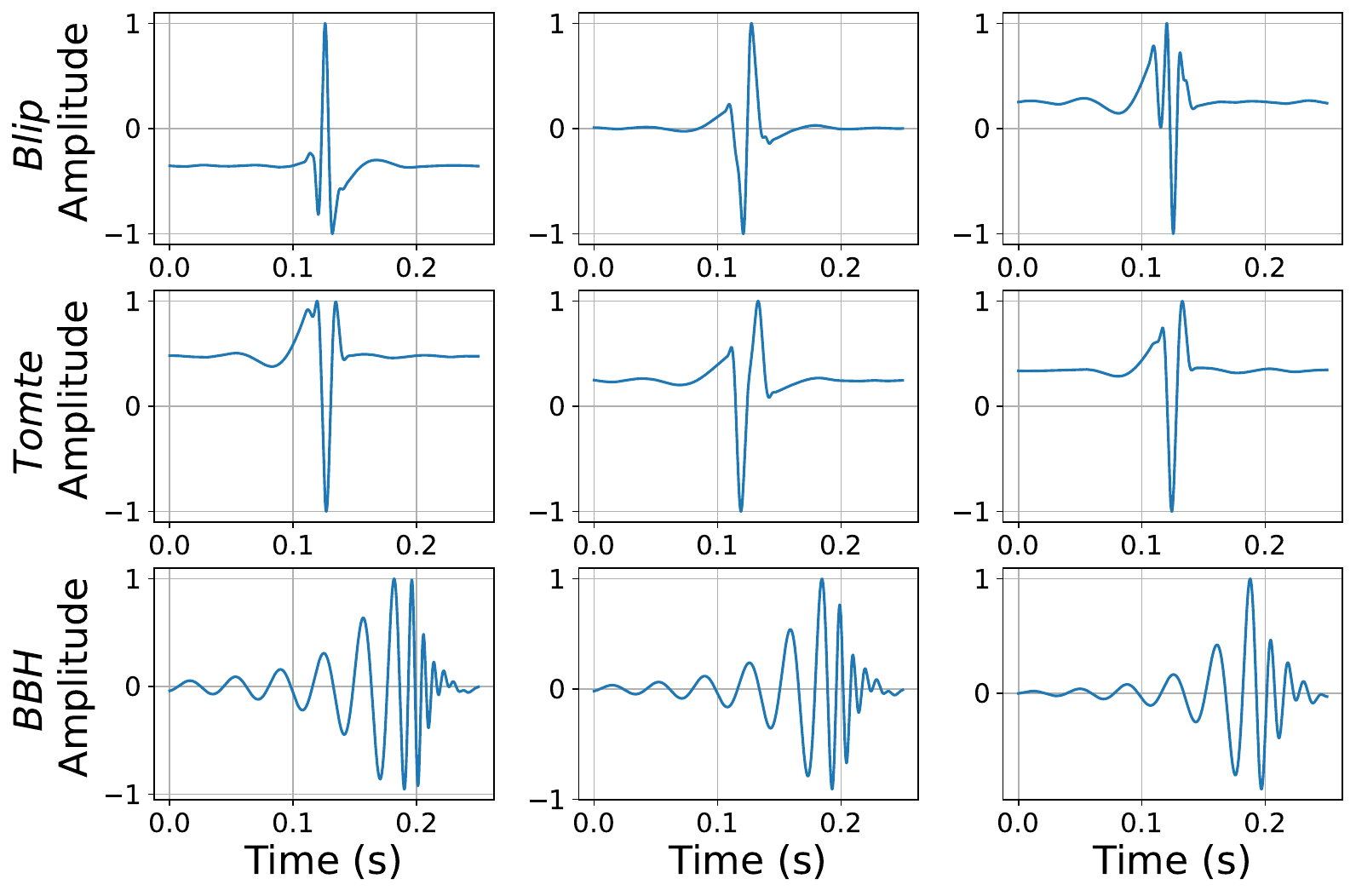}
\caption{Examples of blip (top), tomte (middle) and BBH signals (bottom) used to train GAN models}
\label{fig:example_dataset}
\end{figure}

The GAN models in this study are conditioned on three different classes; two classes that are derived from \textit{Gravity Spy} glitch classes called blip and tomte, and a third BBH signal class.
The astrophysical nature that the BBH signals represent is mainly ignored during experimentation, and they are generally treated as another class of transient, `glitch-like' time-series for experimental purposes.
Examples of the three classes can be observed in Figure \ref{fig:example_dataset}.
Blip glitches have a characteristic time-frequency morphology of a symmetric teardrop shape in the range $[30, 500]\,$Hz with short-durations ($\sim 0.04\,$s). 
They appear in LIGO Livingston and LIGO Hanford, Virgo and GEO $600$ \cite{Cabero:2019orq}.
Due to their abundance and form, they hinder both the unmodeled burst and modelled CBC searches, with particular emphasis on compact binaries with large total mass, highly asymmetric component masses, and spins anti-aligned with the orbital angular momentum  \cite{abbott2018effects, abbott2016characterization}.
Tomte glitches are also short-duration ($\sim 0.25\,$s) with characteristic triangular morphology. 
Since both blip and tomte glitches have no clear correlation to the auxiliary channels, they cannot be removed from astrophysical searches.
The BBH class represents the inspiral and merger of a binary black hole system.
All samples are of length $1,024$ and have a sampling rate of $4,096\,$Hz, corresponding to $0.25\,$s of data. 
The GANs are trained on $7,500$ samples ($2,500$ samples from each class).

The blip and tomte datasets are constructed using confidences from \textit{Gravity Spy} applied to the glitch triggers in LIGO's third observing run (O3) \cite{gspy_data_quality}. 
Only blip and tomte events with \textit{Gravity Spy} confidences of $c^1_{GS} \geq 0.9$ for their respective class are used in this study, and are extracted using \textit{GWpy}'s \cite{gwpy} \textit{fetch\_open\_data} method, which provides an interface to the \textit{GWOSC}\footnote{\href{https://gwosc.org/}{The Gravitational Wave Open Science Centre}} \cite{open_data} data archive.
The glitches are surrounded by stationary and uncorrelated noise, which would hinder the learning of GAN models. 

To avoid the computational expense of \textit{BayesWave}, as done in \cite{GENGLI}, the glitches are isolated from the background using Savitsky-Golay \cite{savitzsky} filters.
This requires the following preprocessing steps:

\begin{enumerate}
    \item Firstly, $20\,$s of strain data is extracted, centred around the glitch GPS time provided by Gravity Spy.
    \item The data is whitened, and a bandpass filter between $(20, 350)\,$Hz is applied to the $20\,$s of data.
    \item The data is cropped at 8,192 datapoints, centred around the GPS time of the glitch, corresponding to $2\,$s of data.
    \item A window of 100 data points is isolated around the centre of the glitch, and two consecutive Savitsky-Golay filters with a polynomial of order 3 are applied to each side around the glitch centre iteratively, with window sizes of 501 and 301 respectively. 
    \item The two smoothed sides and the unsmoothed central peak are then concatenated, followed by applying 3 additional Savitsky-Golay filters iteratively with window sizes 41, 31 and 21 to the entire concatenated sample. The entire sample is finally cropped at 1,024 data points around the event GPS time, centred around 0 and rescaled to $(-1,1)$. 
\end{enumerate}

\begin{figure}[t!]
\centering
\captionsetup{width = 0.45 \textwidth}
\begin{subfigure}[t]{0.172\textwidth}
  \centering
  \includegraphics[width = \textwidth]{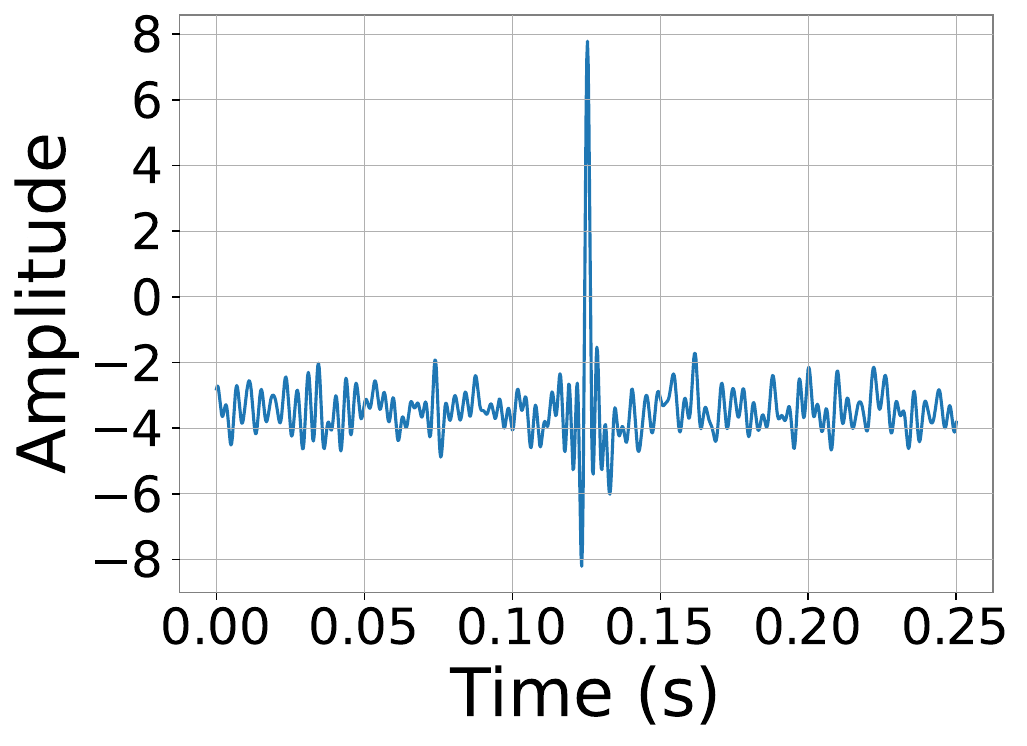}
  \caption{Whitened and bandpassed sample}
  \label{fig:Blip_1}
\end{subfigure}%
\begin{subfigure}[t]{0.16\textwidth}
  \centering
  \includegraphics[width = \textwidth]{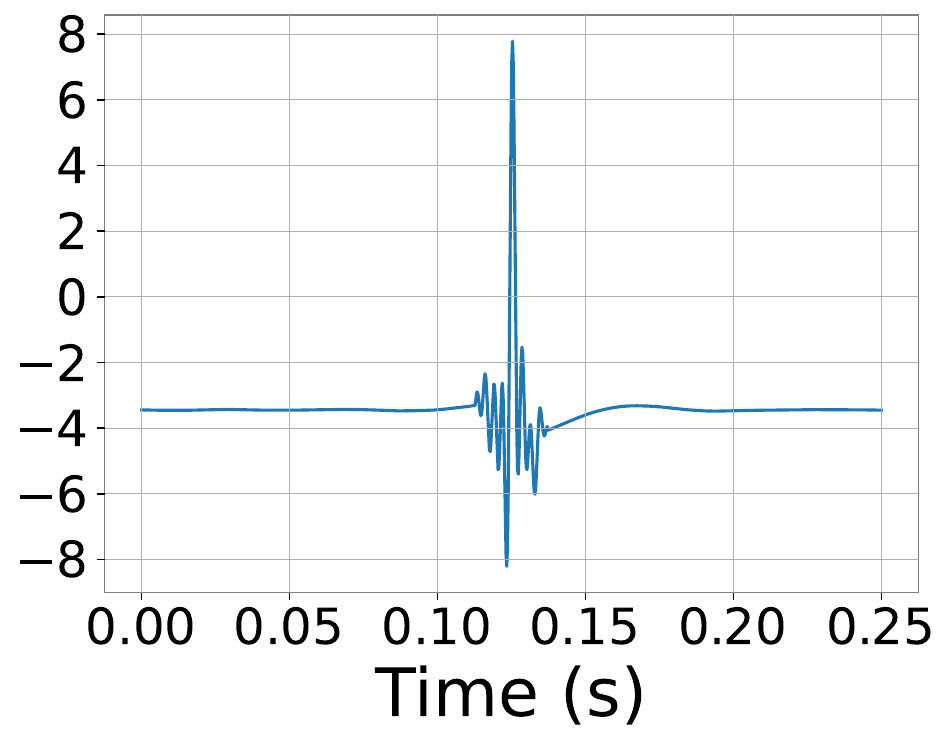}
  \caption{Filtering either side of the peak}
  \label{fig:Blip_2}
\end{subfigure}%
\begin{subfigure}[t]{0.16\textwidth}
  \centering
  \includegraphics[width = \textwidth]{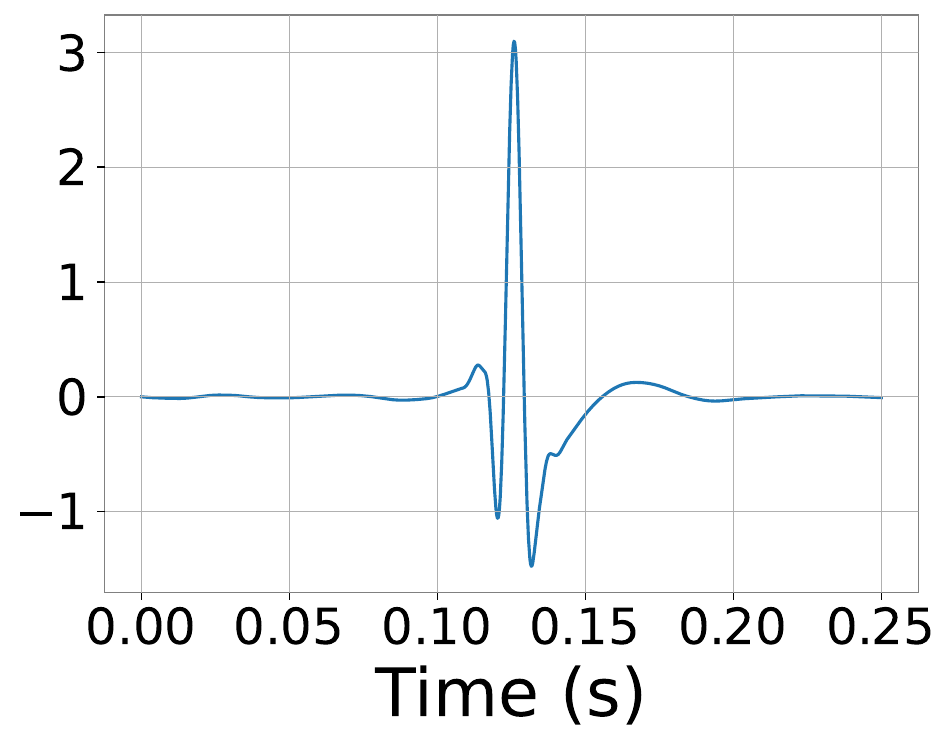}
  \caption{A fully preprocessed blip (before rescaling)}
  \label{fig:Blip_3}
\end{subfigure}%
\caption{Visualizations of the preprocessing steps applied to a blip glitch event. 
}
\label{fig:Preprocess_blip_ex}
\end{figure}

The results of the above preprocessing steps are shown in Figure \ref{fig:Preprocess_blip_ex}. 
The extensive use of Savitsky-Golay filters ensures that high-frequency noise artefacts are removed while preserving the overall shape.
An analysis was made to investigate whether the characteristics of blips and tomtes are preserved after filtering (Appendix \ref{sec:Gspy_analysis}). 
\textit{Gravity Spy} generally classifies the preprocessed samples as their correct class when embedded in detector background for the signal-to-noise ratio (SNR) ranges considered in experiments (see section \ref{sec:Experimental_Procedure}), indicating that their morphologies are analogous to glitches.

\begin{figure}
\centering
\captionsetup{width = 0.45 \textwidth}
  \includegraphics[width=0.4\textwidth]{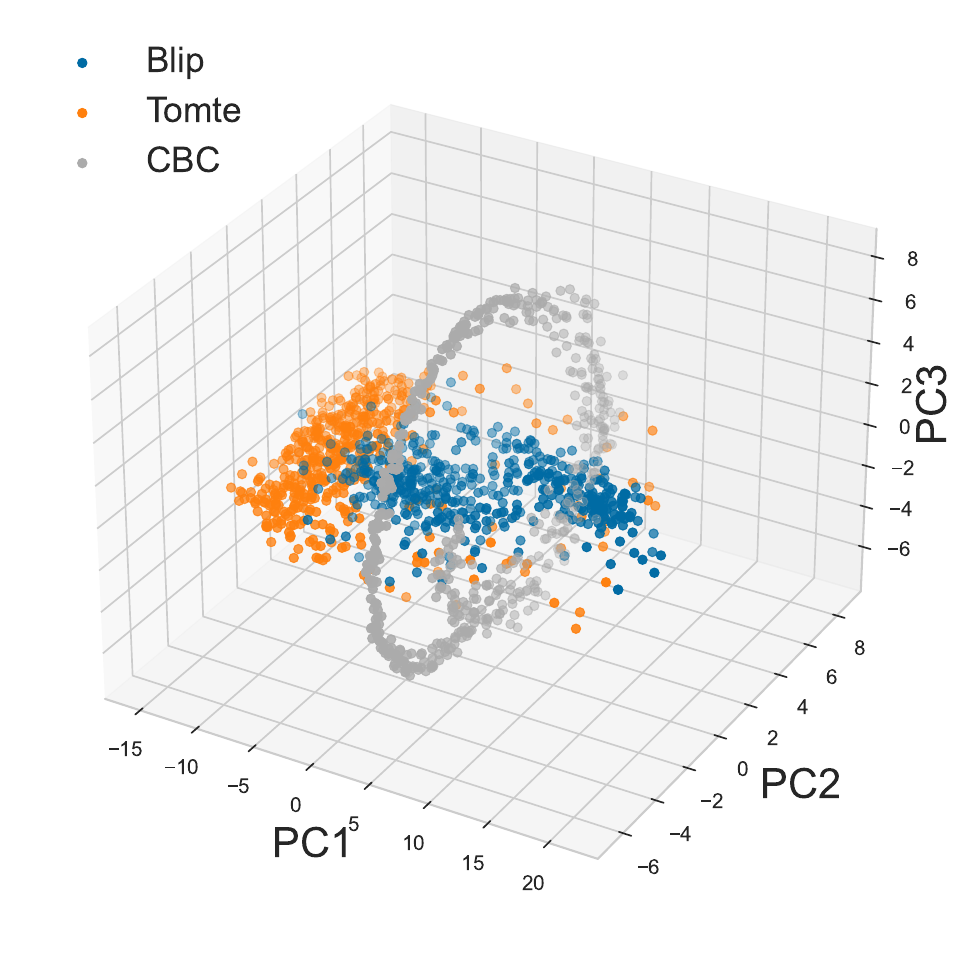}
\caption{A plot of the first 3 principal components of the original samples. The separability of the three classes in this compressed representation indicates the diversity of the data.}
\label{fig:PCA_signals}
\end{figure}
 
All BBH signals are simulated with \textit{PyCBC} \cite{PYCBC_cite} using the IMRPhenomD waveform routine from \textit{LALSuite} \cite{lalsuite}, which generates the inspiral, merger and ringdown of a BBH waveform. 
The component masses are restricted to the range of $[30, 160]M_{\odot}$ with a spin of zero and fixing $m_1>m_2$ and using only plus polarization. 

We show the diversity of the three classes in Figure \ref{fig:PCA_signals}, where the dataset is represented by three separable clusters in a reduced principal component space.

\subsection{\label{sec:Experimental_Procedure}Experimental Procedure}

\subsubsection{GAN Benchmarks\label{sec:GAN_benchmarks}}

This work uses ablation studies to compare cDVGAN and its cDVGAN2 extension with three other baseline GAN models in their ability to generate data useful for training detection algorithms (see section \ref{sec:CNN_classifier}). 
An ablation study in machine learning is a systematic experimentation technique used to understand the contribution of individual components of a model to its overall performance.
In this case, it involves selectively disabling auxiliary first and second-order derivative discriminators during the training of GAN models to investigate if they are successful in improving the features captured in synthetic GW signals and glitches.
To construct an appropriate ablation study, the first benchmark is a conditional variant of a vanilla Wasserstein GAN (cWGAN), which comprises the same architecture as cDVGAN except for the derivative discriminator.
This allows us to investigate the effect of including a first-order derivative discriminator during GAN training.
cDVGAN2 allows us to investigate the effect of including a second-order derivative discriminator on top of the first-order derivative discriminator.

The second benchmark is McGANn \cite{McGinn_2021} since it successfully replicates the features of simulated GW bursts.
As a third benchmark, we developed a modified McGANn model, called McDVGANn, that uses a second auxiliary discriminator applied to first-order derivatives, similar to cDVGAN. 
The architecture is identical to McGANn except for the addition of a derivative discriminator. In McDVGANn, the derivative discriminator is identical to the base discriminator except for the input size.
This benchmark investigates whether the idea of derivative discriminators can generalize to other GAN models.
For a more detailed description of the baselines (see Appendix \ref{sec:McDVGANn_1}).

McGANn and McDVGANn are conditioned using concatenation similar to their paper \cite{McGinn_2021} and are trained with an Adam optimizer, binary cross-entropy loss function and learning rate $2\times10^{-4}$. 
Conversely, cWGAN and cDVGAN2 are conditioned via projection similarly to cDVGAN and are trained with RMSProp, a Wasserstein loss function and a learning rate of $2\times10^{-4}$. 
All models are trained for 500 epochs with a batch size of 512.
Unless otherwise specified, all GANs are trained using the same standard hyperparameters as prior works. 

\subsubsection{GAN-generated datasets\label{sec:GAN_datasets}}

We construct 3 different datasets from each GAN by sampling the respective generator's class space, as suggested by \cite{McGinn_2021}. The three variants of GAN-generated datasets are as follows:
\begin{enumerate}
    \item \textbf{Vertex:}  The vertex class space corresponds to the vertices of the three-dimensional class space. The locations of the vertex class space are the same as the training set class space locations used to train the GAN models and are the closest representation to the training set. The vertex class vector is one-hot encoded corresponding to one of the three GAN training classes eg. [1, 0, 0] corresponds to the blip class.
    \item \textbf{Simplex:} The simplex class space corresponds to points on a $k=2$ simplex (2D triangle) in the case of three classes. Class vectors are constructed by sampling points uniformly on this simplex. The simplex can be considered the simplest surface that intersects all three training classes, and all simplex class vectors sum to 1. Variations are observed in the samples, with some having characteristics that strongly resemble the training classes, due to one class dominating the others. The simplex dataset is a superset of the vertex dataset and represents synthetic data outside the training data distribution.
    \item \textbf{Uniform:} For the uniform dataset, each entry in the class vector is uniformly sampled from U$[0, 1]$, corresponding to sampling uniformly within a cube with dimensions 1x1x1. The uniform dataset is a superset of the simplex and vertex datasets and, like the simplex dataset, represents data from outside of the training data distribution. The uniform dataset exhibits the largest variety since it explores regions of the class space further than the simplex dataset relative to the training set vertices. 
\end{enumerate}

Figure \ref{fig:mixed_generations} shows examples from each of the cDVGAN-generated datasets (also in Appendix \ref{sec:Standard_class_hybrid}). 
Figure \ref{fig:PCA_real_vs_fake_analysis} shows plots of the first 3 principal components (PCs) of real and GAN-generated samples from cDVGAN.
The vertex samples (Figure \ref{fig:PCA_real_vs_fake_vertex}) generally match their corresponding classes from the real data distribution. Figure \ref{fig:PCA_real_vs_fake_all} shows that the simplex and uniform hybrid datasets populate spaces between the class clusters, while Figure \ref{fig:PCA_hybrid} shows that the uniform dataset covers a larger part of the PC space than the simplex dataset. This is intuitive since the simplex space is a subset of the uniform space.

\begin{figure}[ht!]
\centering
\captionsetup{width = 0.5 \textwidth}
\includegraphics[width = 0.5\textwidth]{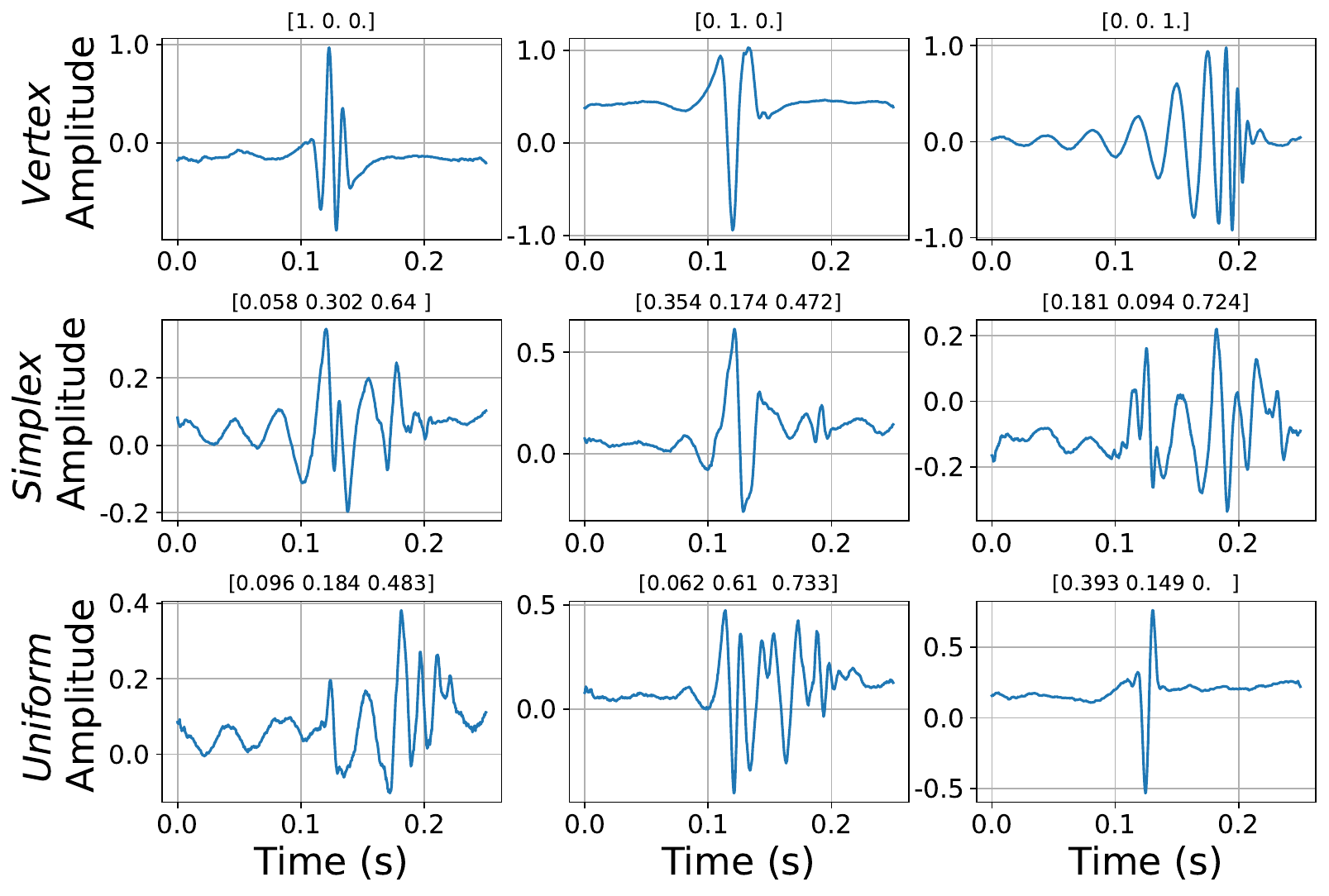}
\caption{Examples of vertex (top), simplex (middle) and uniform samples (bottom) from cDVGAN. The corresponding class vector is shown above each sample.}
\label{fig:mixed_generations}
\end{figure}

\begin{figure*}[t!]
\centering
\captionsetup{width = 0.9 \textwidth}
\begin{subfigure}[t]{0.32\textwidth}
  \centering
  \includegraphics[width = \textwidth]{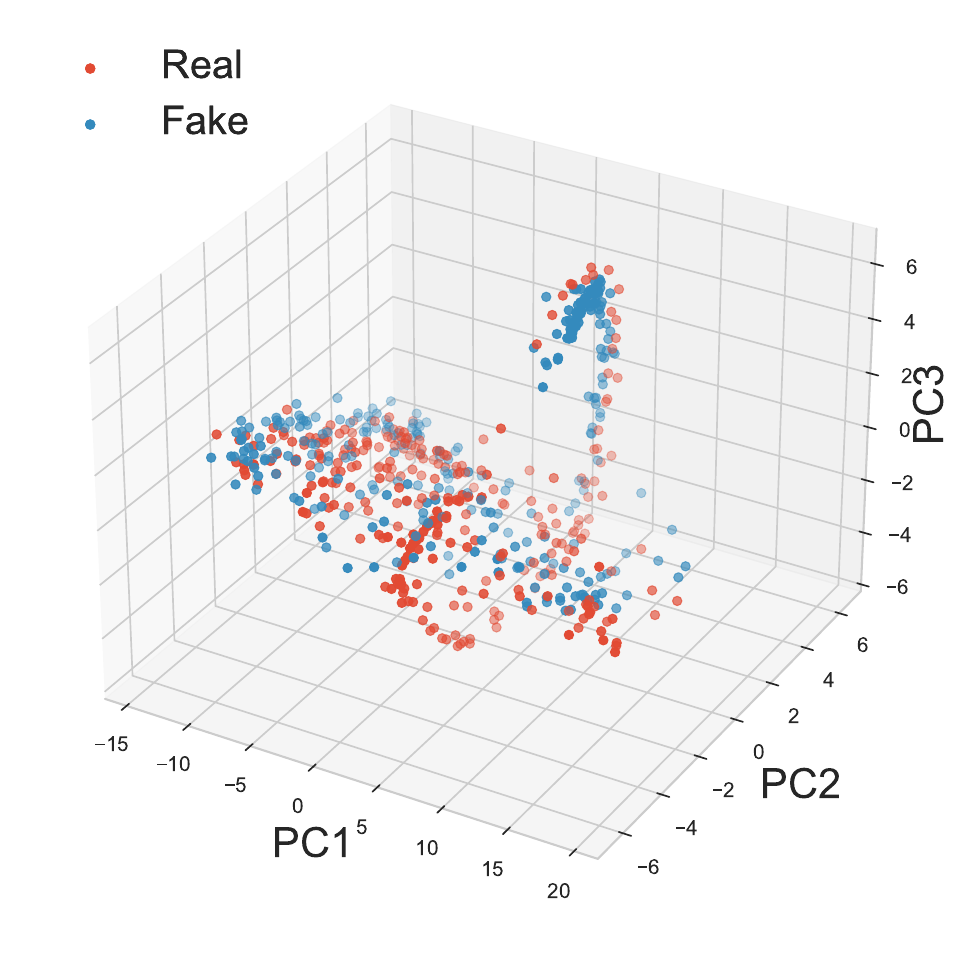}
  \caption{Real and Vertex datasets}
  \label{fig:PCA_real_vs_fake_vertex}
\end{subfigure}%
\begin{subfigure}[t]{0.32\textwidth}
  \centering
  \includegraphics[width = \textwidth]{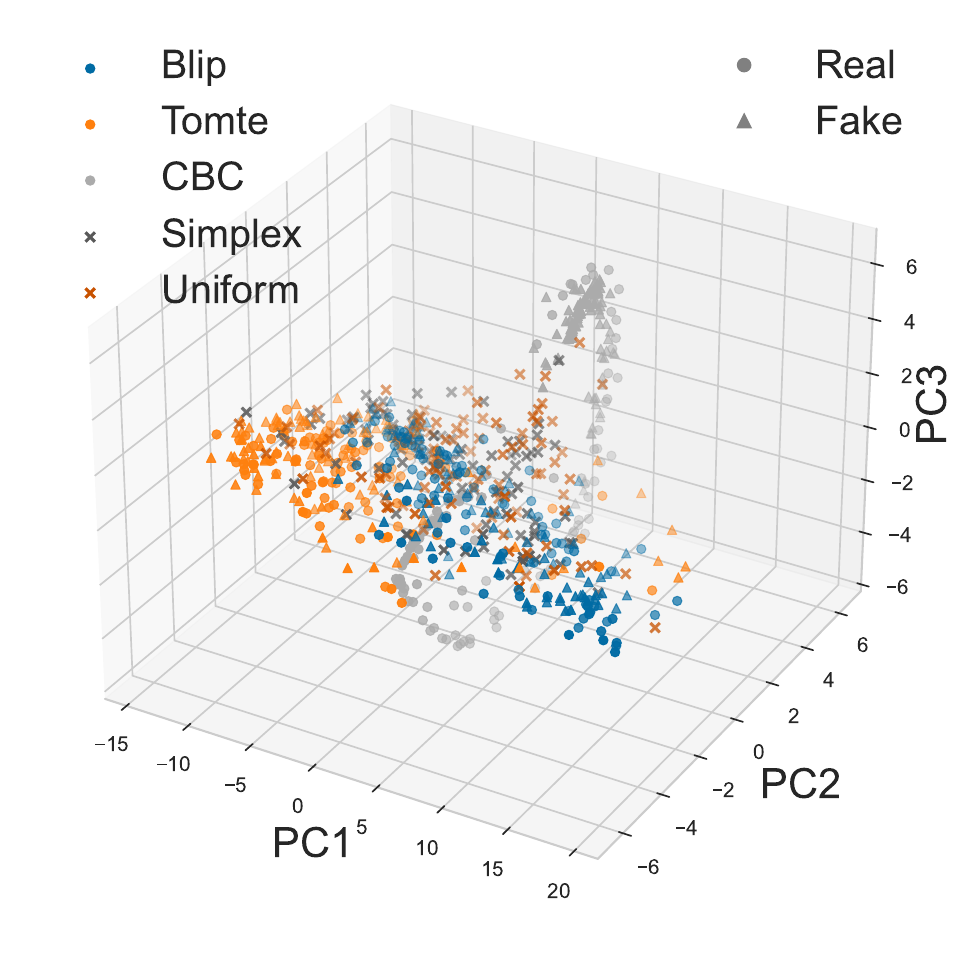}
  \caption{All datasets}
  \label{fig:PCA_real_vs_fake_all}
\end{subfigure}%
\begin{subfigure}[t]{0.32\textwidth}
  \centering
  \includegraphics[width = \textwidth]{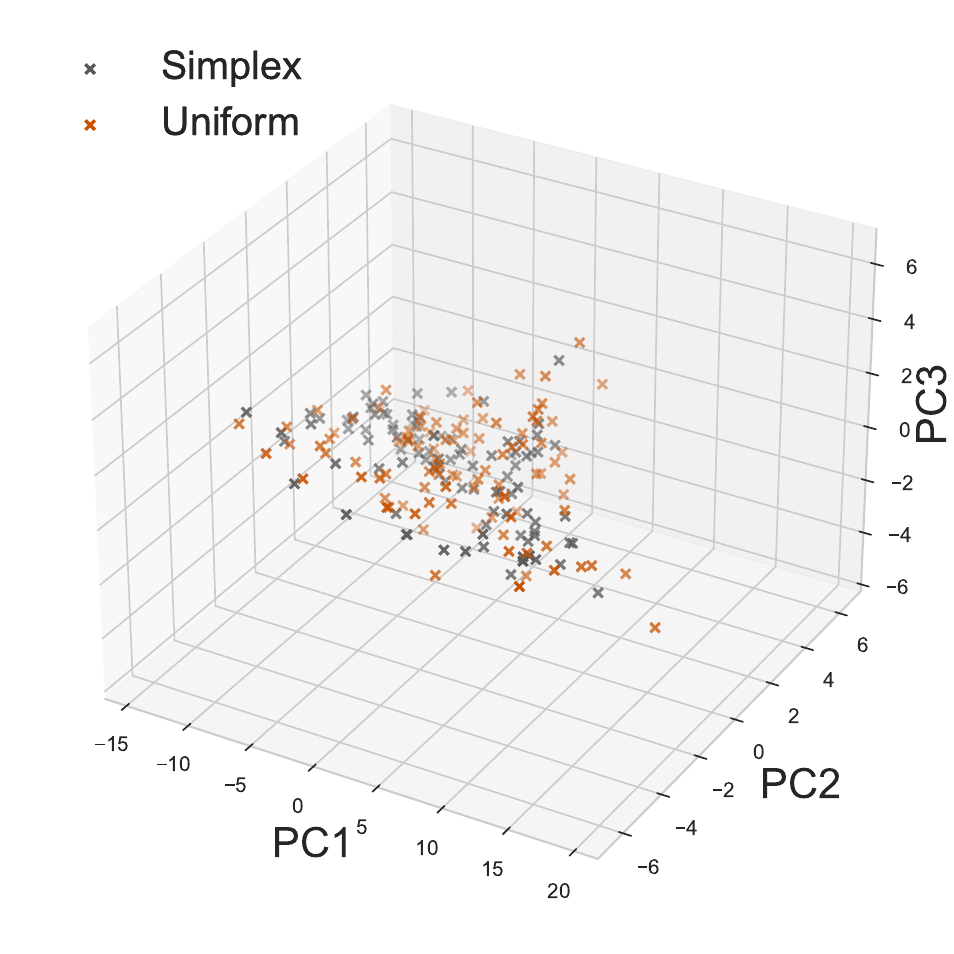}
  \caption{Hybrid datasets}
  \label{fig:PCA_hybrid}
\end{subfigure}%
\caption{The first 3 principal components (PCs) of real and GAN-generated samples from cDVGAN. The vertex samples from cDVGAN generally match the real samples in the PC space while hybrid samples populate intermediate regions between the clusters for the 3 classes. Figure \ref{fig:PCA_hybrid} shows that the uniform dataset covers a larger space than the simplex dataset.}
\label{fig:PCA_real_vs_fake_analysis}
\end{figure*}

\subsubsection{Downstream search with CNNs\label{sec:CNN_classifier}}

The experiments followed in this study investigate the effectiveness of GAN-generated data for training CNNs to detect real data in additive Gaussian noise.
CNNs are a class of deep learning model commonly used in computer vision. 
However, they can also be applied to other spatially adjacent data types like time-series \cite{CNN_TS} and have been applied to GW detector strain data to detect merging black holes \cite{ML_GW_2}.

\begin{figure}[t!]
\centering
\captionsetup{width = 0.45 \textwidth}
\begin{subfigure}[t]{0.25\textwidth}
  \centering
  \includegraphics[width = \textwidth]{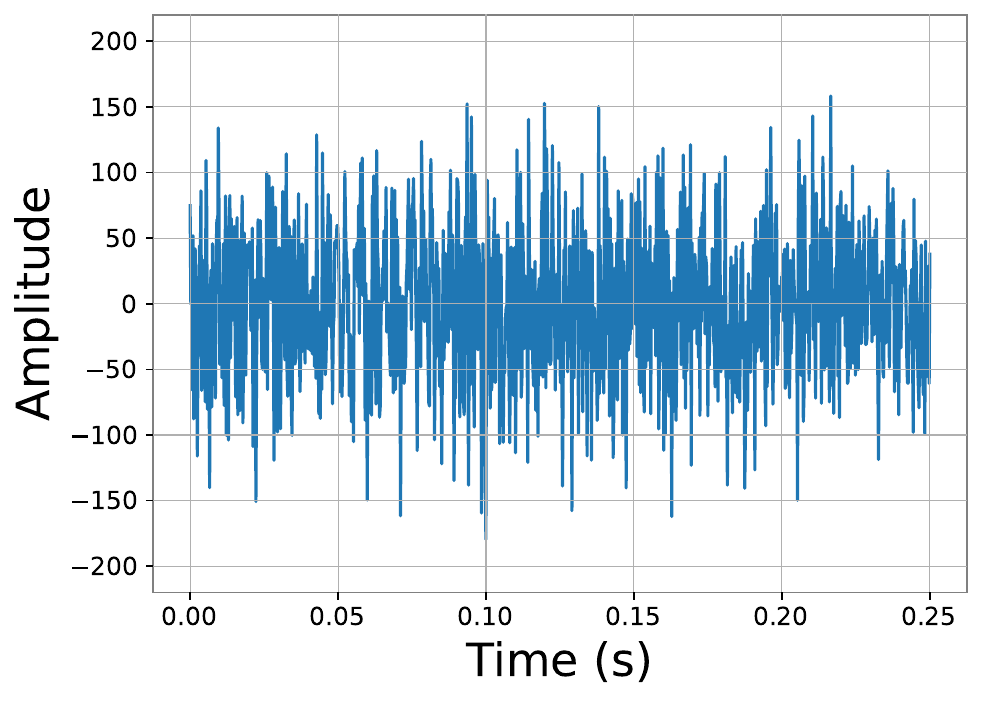}
  \caption{Pure LIGO detector noise (0 class)}
  \label{fig:injected_blip}
\end{subfigure}%
\begin{subfigure}[t]{0.25\textwidth}
  \centering
  \includegraphics[width = \textwidth]{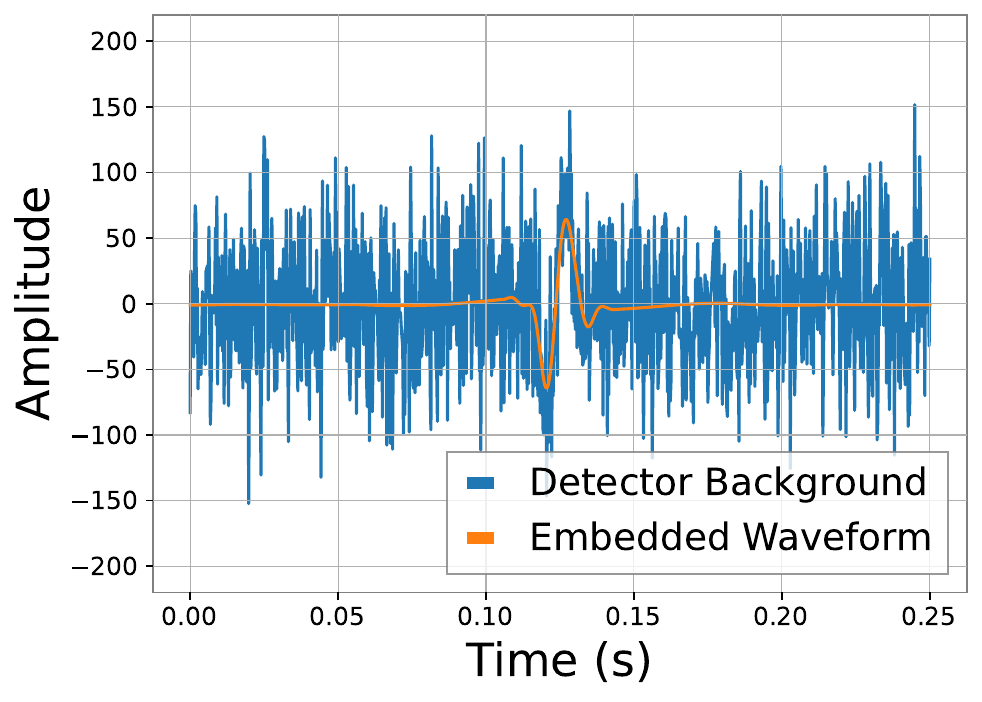}
  \caption{Detector noise + glitch \ \ \ (1 class)}
  \label{fig:non_injected}
\end{subfigure}%
\caption{Examples of the two classes predicted by CNN models. The injected sample on the right is scaled to an SNR of 8 before injection (shown in orange for clarity).
}
\label{fig:injection_examples}
\end{figure}

The CNNs' objective is to perform the binary classification of two classes: samples in additive detector noise and detector noise only, examples of which are shown in Figure \ref{fig:injection_examples}.
It takes a time-series input of dimension $1,024$, representing $0.25\,$s of LIGO strain data. 
The CNN architecture is kept constant for training with each dataset and can be viewed in the Appendix (\ref{sec:CNN_Architecture}). 

Better synthetic data will result in CNNs with better detection efficiency on a real data subset from the GAN training data distribution.
We also investigate if training CNNs with GAN-generated hybrid datasets (simplex and uniform) can improve the detection efficiency beyond training solely with the standard three classes.

The training and testing samples are injected additively into detector noise from Hanford (H1) or Livingston detectors (L1) during O3 for each of the above datasets.
The detector noise is extracted from \textit{GWOSC}, using trigger times for \textit{Gravity Spy's} \textit{No\_glitch} class, extracing $14\,$s around each \textit{No\_glitch} GPS time.
Although it is not investigated whether other glitches are present in the noise, using \textit{Gravity Spy's} \textit{No\_glitch} class should guarantee that most of the samples contain stationary noise after whitening. 

The background is sampled at $4,096\,$Hz, similar to the GAN training data.
The LIGO detector noise is first whitened using \textit{PyCBC}'s whitening function, with the first and last $2.5\,$s removed due to artefacts at the noise boundaries.
The preprocessed detector noise is then split into chunks of length $1,024$, the same dimensionality as the GAN input and output.
Following this procedure, we accumulate just over $250,000$ background samples in total.
Since the GAN training data and output are scaled between $[-1, 1]$ the samples are scaled to a signal-to-noise ratio (SNR) ratio that is sampled uniformly on U$[1, 16]$ before injecting them into the preprocessed detector noise.
This is done by first computing $\rho_{opt}$ for each generated sample according to

\begin{equation}\label{eq:snr_scaling}
    \rho_{opt}^2 = 4\int^{f_{max}}_{f_{min}}\frac{|\hat{h}(f)|^2}{S_n(f)}df
\end{equation}

where $\hat{h}(f)$ and $S_n(f)$ are the Fourier transform of the input sample (blip, tomte, BBH) and the detector noise power spectral density (PSD) respectively \cite{Allen_2012_findchirp} (which we set to unity for convenience as we are working with whitened data). 
The sample can then be scaled to the desired $\rho_{opt}$ on U$[1, 16]$.

For each GAN, we train three separate CNNs on the three generated datasets (vertex, simplex, uniform).
For the vertex dataset the three different vertex locations in the class space are sampled with equal probability.
For the uniform and simplex datasets, samples are drawn uniformly from their respective spaces. 

\begin{figure*}[ht]
\centering
\captionsetup{width = 0.9 \textwidth}
\begin{subfigure}{0.18\textwidth}
  \centering
  \includegraphics[width = \textwidth]{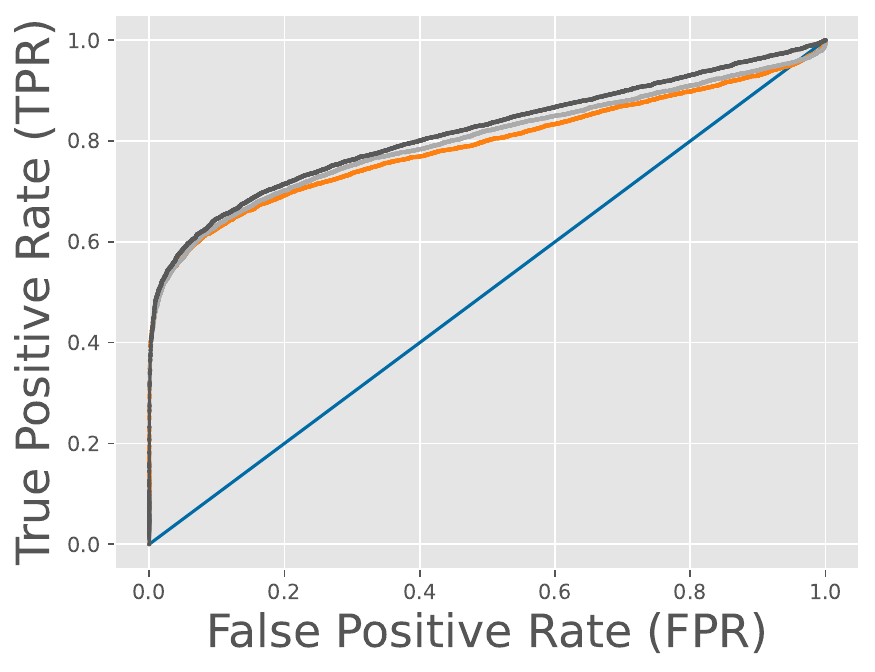}
  \caption{cDVGAN}
  \label{fig:result_DVGAN_roc
  }
\end{subfigure}%
\begin{subfigure}{0.18\textwidth}
  \centering
  \includegraphics[width = \textwidth]{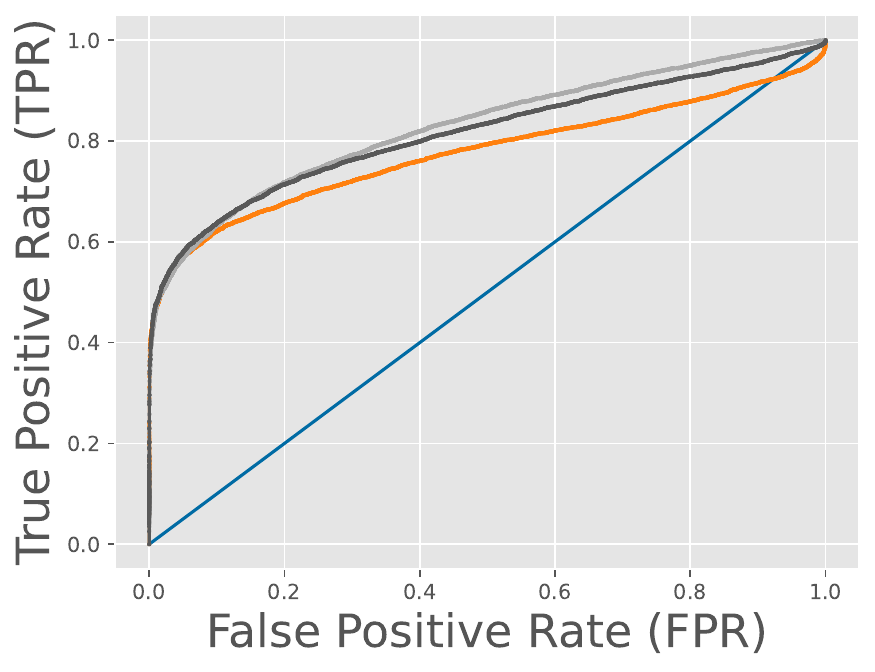}
  \caption{cDVGAN2}
  \label{fig:result_DVGAN2_roc}
\end{subfigure}%
\begin{subfigure}{0.18\textwidth}
  \centering
  \includegraphics[width = \textwidth]{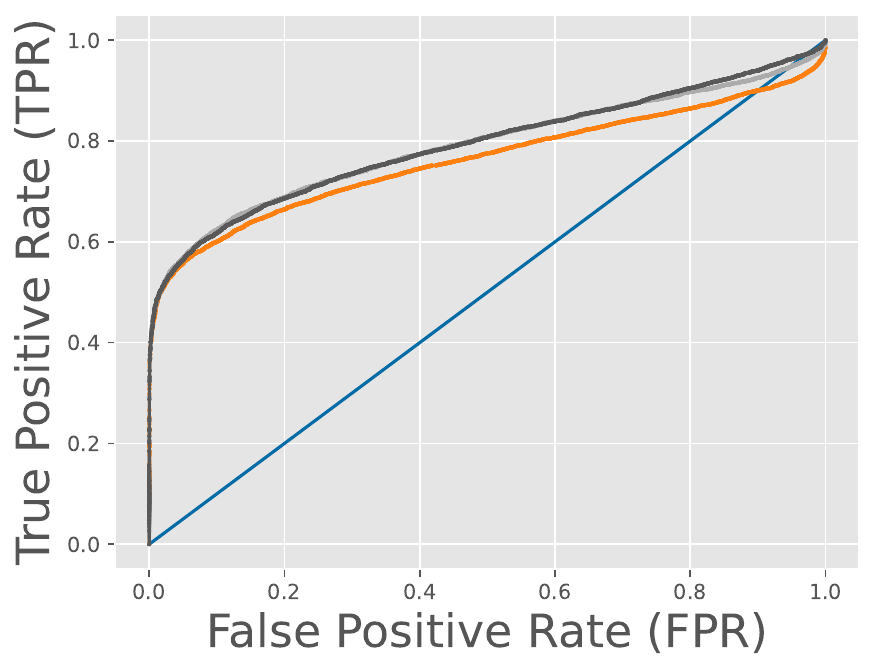}
  \caption{cWGAN}
  \label{fig:result_WGAN_roc}
\end{subfigure}%
\begin{subfigure}{0.18\textwidth}
  \centering
  \includegraphics[width = \textwidth]{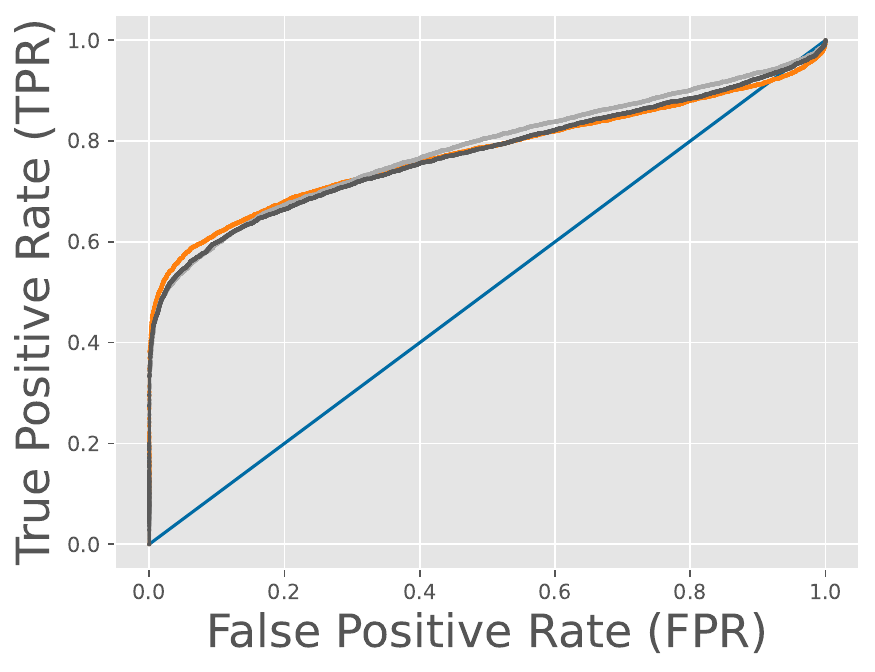}
  \caption{McGANn}
  \label{fig:result_MCGANN_roc}
\end{subfigure}%
\begin{subfigure}{0.18\textwidth}
  \centering
  \includegraphics[width = \textwidth]{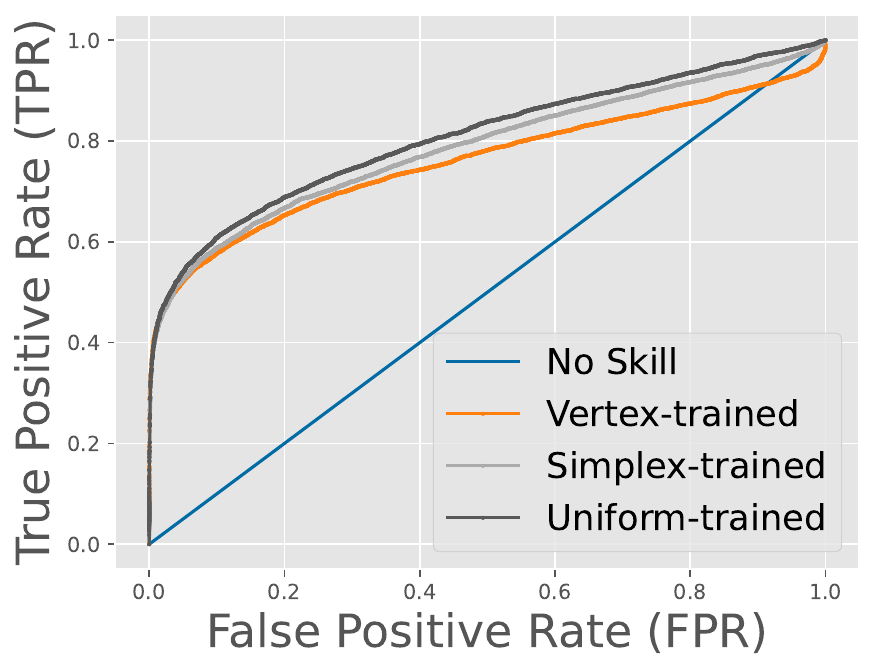}
  \caption{McDVGANn}
  \label{fig:result_MCDVGANN_roc}
\end{subfigure}%
\caption{ROC-Curves for the different GAN-generated datasets from each GAN.}
\label{fig:results_gans_rocs}
\end{figure*}

\begin{table*}[ht]
\centering
\begin{tabular}{l|l|l|l|l|l} 
    \hline
    {\textbf{Dataset}} & {\textbf{cDVGAN (ours)}} & \textbf{cDVGAN2 (ours)} & {\textbf{cWGAN}} & {\textbf{McGANn}} & \textbf{McDVGANn (ours)} \\ [0.1cm] \hline 
    \textbf{Vertex-Trained} & 0.771 $\pm$ 0.012 & 0.758 $\pm$ 0.008 & 0.762 $\pm$ 0.018 & 0.768 $\pm$ 0.016 & 0.768 $\pm$ 0.022  \\ [0.2cm]
    \textbf{Simplex-Trained} & \textit{\textbf{0.802}} $\pm$ 0.019  & 0.789 $\pm$ 0.010  & \textit{0.788} $\pm$ 0.009 & 0.759 $\pm$ 0.010 & \textit{0.786} $\pm$ 0.012   \\ [0.2cm]
    \textbf{Uniform-Trained} & 0.797 $\pm$ 0.022 & \textit{0.791} $\pm$ 0.009 & 0.777 $\pm$ 0.014 & \textit{0.770} $\pm$ 0.014 & 0.778 $\pm$ 0.012\\ [0.1cm]
    \hline \bottomrule
\end{tabular}
\captionsetup{width = \textwidth}
\caption{The Area-Under-Curve (AUC) yielded on a real test set by CNNs trained on each synthetic dataset from each GAN. The results represent the mean AUC over 5 iterations, where the bounds are calculated using the standard deviations over the 5 iterations. The best result overall is shown in bold text, while the best result per GAN is shown in italic text.}
\label{tab:AUC_cnns}
\end{table*}

Each training dataset comprises $100,000$ samples, with 50\% glitch/signal plus Gaussian noise and 50\% Gaussian noise only.
We test each CNN on a real data subset comprising $7,500$ samples ($2,500$ from each class), sampled randomly from a distribution of $20,850$ samples ($6,950$ from each class).
This results in a test dataset of 15,000 samples in total (50\% LIGO noise only) in each iteration of training/testing.
We ensure that none of the GAN training data appear in the test set, although they are taken from the same distribution.

The results are presented using the area-under-the-curve (AUC) metric.
AUC is a metric commonly used in machine learning to evaluate the performance of a classification model, particularly in the context of binary classification problems. 
It is associated with the Receiver Operating Characteristic (ROC) curve (see Figure \ref{fig:results_gans_rocs}), which is a graphical representation of the trade-off between true positive rate (sensitivity) and false positive rate (1-specificity) across different thresholds.
It provides a single metric between 0 and 1 that summarizes the overall discriminatory power of a model across different classification thresholds.
We repeat the experiment 5 times and report the mean results, randomly generating the training data and randomly sampling the backgrounds and test data in each iteration.
We keep the real test datasets constant for each GAN-generated dataset in each iteration.


\subsubsection{Fitting-factor study \label{sec:Matched_filter_exp}}
We investigate the accuracy of cDVGAN's BBH signals under a fitting-factor study, showing that most of the synthetic signals are consistent with the original signals simulated with \textit{IMRPhenomD}.
The fitting-factor of a signal is defined as the maximum match of that signal over the templates of a template bank \cite{template_bank_cite}.

While cDVGAN's signals are used as the signal injections in a matched-filter search, the template bank is created using \textit{mbank} \cite{Stefano_mbank}, and the corresponding templates are simulated using \textit{IMRPhenomD}.
We aim for good coverage of the parameter space of cDVGAN's training signals ($30 \leq m2 \leq m1 \leq 160$) beyond the 97\% requirement for matched-filter standards, using a holdout dataset from the cDVGAN BBH distribution to validate the template bank\footnote{We validate the bank using a flat PSD, required for time-domain match calculations using \textit{mbank}}.
We sample \textit{mbank's} normalizing flow until the parameter space is covered evenly with an average fitting factor of over 99\% (only approximately 5\% of signals yield a fitting factor of under 99\%.
This yields $1,500$ templates, allowing for a thorough exploration of the parameter space.

We calculate the fitting factor in the time-domain on a fixed time grid since the training signals and generated signals are truncated at $0.25\,$s around the merger time. We compute the fitting-factor over the $2,500$ GAN training samples simulated using \textit{IMRPhenomD} and compare it to the fitting-factor computed on $2,500$ synthetic signals from cDVGAN.

An important caveat to highlight is that the class-conditional approach featured in cDVGAN is focused towards a glitch generator application rather than a waveform generator for GW searches. 
To achieve better accuracy in the latter case, it would be better to modify the cDVGAN architecture to learn BBH signals only.
This would simplify the training schedule to learn one distribution of signals rather than three distributions in one model. 
Furthermore, the model could also be conditioned on continuous source parameters rather than discrete class information, giving the user control over the parameter space they wish to generate signals in.

\section{\label{sec:Results}Results}
\subsection{\label{sec:Results_GAN}Training with GAN data}

\begin{table*}
\centering
\begin{tabular}{l|l|l|l|l|l} 
    \hline
    {\textbf{Dataset}} & {\textbf{cDVGAN} (ours)} & \textbf{cDVGAN2 (ours)} & {\textbf{cWGAN}} & {\textbf{McGANn}} & \textbf{McDVGANn (ours)} \\ [0.1cm] \hline 
    \textbf{Vertex-Trained} & 0.689 $\pm$ 0.009 & 0.680 $\pm$ 0.006 & 0.685 $\pm$ 0.017 & \textit{0.687} $\pm$ 0.010 & 0.668 $\pm$ 0.020   \\ [0.2cm]
    \textbf{Simplex-Trained} & 0.698 $\pm$ 0.013  & 0.686 $\pm$ 0.007 & \textit{0.695} $\pm$ 0.010 & 0.673 $\pm$ 0.008 & \textit{0.676} $\pm$ 0.010 \\ [0.2cm]
    \textbf{Uniform-Trained} & \textit{\textbf{0.702}} $\pm$ 0.014 & \textit{0.693} $\pm$ 0.008 & 0.692 $\pm$ 0.009 & 0.680 $\pm$ 0.013 & 0.671 $\pm$ 0.006 \\ [0.1cm]
    \hline \bottomrule
\end{tabular}
\captionsetup{width = \textwidth}
\caption{The AUC for test samples under an SNR of 8. The results represent the mean AUC over 5 iterations, where the bounds are calculated using the standard deviations over the 5 iterations. The best result overall is shown in bold text, while the best result per GAN is shown in italic text.}
\label{tab:efficiency_low_snr}
\end{table*}

\begin{table*}
\centering
\begin{tabular}{l|l|l|l|l|l} 
    \hline
    {\textbf{Dataset}} & {\textbf{cDVGAN} (ours)} & \textbf{cDVGAN2 (ours)} & {\textbf{cWGAN}} & {\textbf{McGANn }} & \textbf{McDVGANn (ours)} \\ [0.1cm] \hline 
    \textbf{Vertex-Trained} & 0.844 $\pm$ 0.015 & 0.827 $\pm$ 0.010 & 0.830 $\pm$ 0.022 & 0.841 $\pm$ 0.021 & 0.856 $\pm$ 0.027  \\ [0.2cm]
    \textbf{Simplex-Trained} & \textit{\textbf{0.893}} $\pm$ 0.029  & \textit{0.881} $\pm$ 0.013  & \textit{0.867} $\pm$ 0.010 & 0.836 $\pm$ 0.012 & \textit{0.885} $\pm$ 0.013  \\ [0.2cm]
    \textbf{Uniform-Trained} & 0.883 $\pm$ 0.031 & 0.878 $\pm$ 0.011 & 0.851 $\pm$ 0.020 & \textit{0.852} $\pm$ 0.020 & 0.873 $\pm$ 0.014 \\ [0.1cm]
    \hline \bottomrule
\end{tabular}
\captionsetup{width = \textwidth}
\caption{The AUC for samples above an SNR of 8. The results represent the mean AUC over 5 iterations, where the bounds are calculated using the standard deviations over the 5 iterations. The best result overall is shown in bold text, while the best result per GAN is shown in italic text.}
\label{tab:efficiency_high_snr}
\end{table*}

The AUC values for each CNN model over the entire SNR range (1-16) are shown in Table \ref{tab:AUC_cnns}. 
We also scale the test sets between SNRs of 1-8 and 8-16 and record the AUC of the same CNNs to investigate performance for quieter and louder samples, which can be seen in Tables \ref{tab:efficiency_low_snr} and \ref{tab:efficiency_high_snr} respectively.
Examples of ROC curves from one of the testing iterations are shown in Figure \ref{fig:results_gans_rocs}.
All tables show that the simplex and uniform datasets from cDVGAN yield the highest AUC results for CNNs, with the simplex dataset yielding the highest overall AUC.

\textbf{Ablation Studies.} 
The ablation between cDVGAN and cWGAN in Table \ref{tab:AUC_cnns} shows that the adversarial feedback from the first-order derivative discriminator improves the features of the synthetic data in all three datasets. 
This suggests that the first-order derivative discriminator was successful in improving the GAN output.
cDVGAN2 was unsuccessful in improving upon cDVGAN's results, yielding a slightly lower performance, but improves upon the performance of cWGAN's simplex and uniform datasets, particularly for the higher SNR range (Table \ref{tab:efficiency_high_snr}).

The AUC performance yielded from McDVGANn's datasets is competitive among the GAN models.
Both its simplex and uniform datasets yield better AUC performance than McGANn's counterparts, with the simplex dataset yielding the highest overall AUC performance between the two models.
The vertex datasets from McGANn and McDVGANn yield comparable performance.
This suggests that the derivative discriminator can be effective in a more traditional GAN architecture.

These results show that incorporating derivative discriminators can improve the synthetic data in multiple GAN architectures, and indicate that analyzing first-order derivatives in a separate auxiliary discriminator is superior to using both first and second-order derivative discriminators for modelling the dataset covered in this study (in the cWGAN architecture).
For all GAN models, hybrid datasets provide the best overall AUC performance for the CNN.
This suggests that GAN-generated hybrid samples are useful for searching for multiple classes of real data when obscured by the detector background.
This might offer interesting applications in glitch searches, particularly for those with no clear correlation to auxiliary channels.
Such glitches could be conditioned into cDVGAN to generate hybrid samples specific to a subset of LIGO glitch classes for use in a glitch detection algorithm.
Combining all three GAN-generated datasets for training CNNs may improve upon these results yet again, although this is left to future work.

\begin{table*}
\centering
\begin{tabular}{l|l|l|l|l|l} 
    \hline
    {\textbf{SNR}} & {\textbf{100:0}} & \textbf{75:25} & {\textbf{50:50}} & {\textbf{25:75}} & \textbf{0:100} \\ [0.1cm] \hline 
    \textbf{1-16} & 0.900 $\pm$ 0.001 & 0.898 $\pm$ 0.002 & 0.893 $\pm$ 0.002 & 0.887 $\pm$ 0.002 & 0.802 $\pm$ 0.019  \\ [0.2cm]
    \textbf{1-8} & 0.799 $\pm$ 0.001  & 0.792 $\pm$ 0.005  & 0.787 $\pm$ 0.003 & 0.777 $\pm$ 0.003 & 0.698 $\pm$ 0.013 \\ [0.2cm]
    \textbf{8-16} & 0.988 $\pm$ 0.001 & 0.987 $\pm$ 0.001 & 0.987 $\pm$ 0.001 & 0.985 $\pm$ 0.001 & 0.893 $\pm$ 0.029 \\ [0.1cm]
    \hline \bottomrule
\end{tabular}
\captionsetup{width = \textwidth}
\caption{AUC values over three SNR ranges for different proportions of real:synthetic samples for a training set fixed at 100,000 samples. The results are represented by the mean AUC and standard deviation over 5 iterations.}
\label{tab:real_fake_combined}
\end{table*}

\subsection{\label{sec:result_real}Combining real and cDVGAN data for improved training}
In this section, we augment real datasets with GAN-generated data to improve upon the classification performance in the previous section. 
We augment the real data with the simplex dataset from cDVGAN since it yields the best performance.
We compare cDVGAN data with traditional duplication of real training samples for data augmentation. 

Maintaining a real hold-out test set of 7,500 samples (2,500 from each class) as in section \ref{sec:Results_GAN}, we use all remaining samples from the real distribution for training, which amounts to 13,350 samples (4,450 from each class). 
Fixing the training data size again at 100,000 (50,000 glitch/signal + noise samples, 50,000 noise-only samples), we vary the proportion of real and GAN-generated samples. 
Since there is no limit to generating cDVGAN data, we duplicate the real training data before injection to reach the required number of samples.
This is done to control the effects of the background noise on the training of CNNs, since CNNs are sensitive to the number of different backgrounds seen during training.

The results in Table \ref{tab:real_fake_combined} show that the performance drops only slightly with smaller proportions of real data. The CNN performance remains competitive even with only 25\% of real training samples with only a 1\% drop in overall AUC performance compared to only using real data.
The second and third rows of Table \ref{tab:real_fake_combined} show that this decrease in AUC performance occurs mostly for lower SNR ($<$8) samples.
The decrease in performance for louder samples is minimal ($<$1\% decrease).
This indicates that the synthetic data is competitive for augmenting the training data for CNNs when including a relatively small amount of real data in the training set.
Although there is a more substantial decrease in performance when using 100\% synthetic samples, this might be improved by including other synthetic datasets from cDVGAN in the training schedule.

\subsection{\label{sec:result_match}Fitting-factor results}

\begin{figure}[b!]
\centering
\captionsetup{width = 0.45 \textwidth}
\begin{subfigure}[t]{0.25\textwidth}
  \centering
  \includegraphics[width = \textwidth]{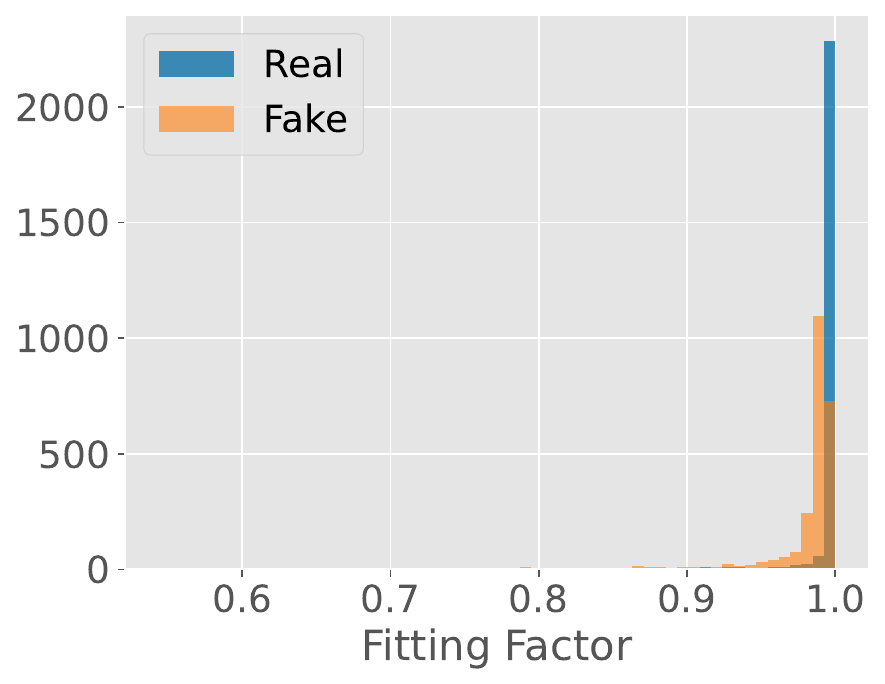}
  \caption{Fitting-factor}
  \label{fig:fitting_factor hist}
\end{subfigure}%
\begin{subfigure}[t]{0.25\textwidth}
  \centering
  \includegraphics[width = \textwidth]{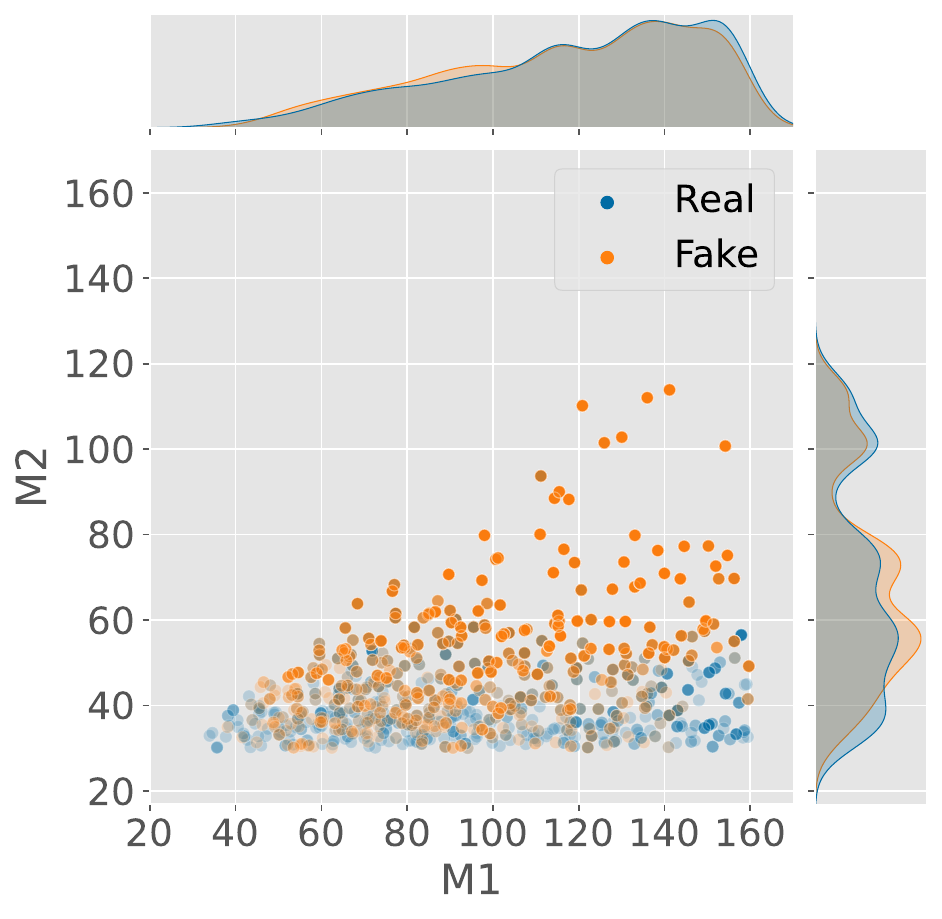}
  \caption{Match parameters}
  \label{fig:Match Parameters}
\end{subfigure}%
\caption{Plots of fitting-factors and corresponding best-fit template parameters for real GAN training signals and synthetic cDVGAN signals.}
\label{fig:ff_study}
\end{figure}

The results indicate that cDVGAN's BBH signals are generally consistent with \textit{IMRPhenomD}.
The search calculates an average fitting-factor of $0.994 \pm 0.0423$ for the $2,500$ GAN training signals with $5^{th}$ and $10^{th}$ percentiles at $0.972$ and $0.994$ respectively.
Conversely, the experiment yields an average fitting-factor of $0.976 \pm 0.045$ for $2,500$ synthetic cDVGAN signals with $5^{th}$ and $10^{th}$ percentiles at $0.893$ and $0.951$ respectively.

Figure \ref{fig:ff_study} shows a histogram of the fitting-factors yielded from the real and synthetic signals and the distribution of the recovered parameters of both datasets.
Figure \ref{fig:fitting_factor hist} shows the fitting-factor distribution of cDVGAN's signals is similar to that of the real training signals, although a minor decrease in accuracy is observed.
Figure \ref{fig:Match Parameters} shows that cDVGAN's signals cover most of the parameter space well, although there is a slight mismatch in the lower ends of both the $M1$ and $M2$ distributions.

Although the accuracy of cDVGAN's signals is lower than that of \textit{IMRPhenomD}, the quality of cDVGAN's signals and coverage of the parameter space could be improved with further training, or by considering a different modelling approach as outlined in Section \ref{sec:Matched_filter_exp}.
Furthermore, cDVGAN is competetive with state-of-the-art surrogate models in terms of inference speed on a CPU \cite{surrogate_model_speed_1}\cite{surrogate_model_speed_2}, and far surpasses them with the use of a GPU.
For example, cDVGAN can generate $2,500$ signals of the learned BBH class in approximately $18 \,$s on a CPU, and only $0.04\,$s on a GPU.

\section{\label{sec:Conclusion}Conclusion}

\subsection{\label{sec:Conclusions}Conclusions}
Time-domain generative modelling in the GAN framework has shown potential to improve GW data analysis. 
Using GANs to learn distributions of GWs and other events of interest such as detector glitches can be useful for data augmentation tasks, validating detection schemes for unmodelled waveforms such as \cite{Omicron, DRAGO2021100678_coherent_pipe, wave_burst}, or be used to construct mock data challenges.
This work presents a novel conditional GAN, called cDVGAN, for generating distinct time-domain classes, including two classes of unmodelled glitches and one class of modelled BBH signals.
cDVGAN uses additional adversarial feedback on the first-order derivatives of training samples in an auxiliary discriminator and generates realistic samples that span the variation within each class.
It also allows for the explicit control of the mixing of classes.
Thus, it is capable of generating generalized hybrid samples that are outside of the limited training distribution and span the variation between classes by sampling the continuous class space.

We use ablation studies to show the effectiveness of using auxiliary discriminators to analyze sample derivatives in an experiment that uses GAN-generated data to train convolutional neural networks (CNNs) to detect real samples in LIGO detector noise.
An ablation study between cDVGAN and its vanilla cWGAN counterpart shows that the additional adversarial feedback from the first-order derivative discriminator yields generated data that is more useful for training a CNN detection algorithm.

The ablation study between cDVGAN2 and cDVGAN reveals that second-order derivative discrimination did not improve the performance under this problem scheme, although the performance of cDVGAN2 is competitive with other baseline GANs.
Another ablation study between our McDVGANn model and McGANn \cite{McGinn_2021} indicates that the method can be effective under a traditional cGAN architecture.
These results suggest that providing adversarial feedback on derivatives on top of the original samples can improve the learning of GANs on continuous time-series, and in particular, events of interest to the GW physics community.

Furthermore, our experiments demonstrate the effectiveness of GAN-generated hybrid samples for training detection algorithms. 
The best overall synthetic dataset for training CNNs was cDVGAN's simplex dataset, while hybrid datasets from other GANs yielded better training sets than the standard vertex dataset. 
We also combine GAN-generated data with real data to improve the performance of CNN models for glitch and signal searches, showing cDVGAN as a viable approach for data augmentation.
Lastly, we implement a fitting-factor study that shows cDVGAN's BBH signals are consistent with the \textit{IMRPhenomD} waveform routine used to generate the cDVGAN training signals.
Although there are some inconsistencies and a small decrease in accuracy, the synthetic signals generally match well with a template bank for the corresponding parameter range. The cDVGAN signals have good coverage on most of the parameter range, and can be generated very efficiently, particularly with the use of a GPU.

Since hyperparameter optimization was not the focus of this research, investigations could be made into better architectures for cDVGAN. 
For example, optimization of the $\eta$ (Equation \ref{eq:combined_generator}) hyperparameters controlling the contribution of each discriminator to the generator loss might yield better generated data.
Including a consistency term, as in \cite{improving_improved_training}, may also improve the generated data from cDVGAN.
Expanding cDVGAN to other representations of the data, might also improve the quality of the generated data.
Finally, research into better CNNs or other detection algorithms that can make use of the GAN-generated data might also result in efficient and scalable analysis solutions towards the next-generation detectors.

Extending cDVGAN to other glitch types is vital to significantly stimulate GW data analysis.
This study takes a step towards this goal, showing how arbitrary time-domain glitches or signals can be conditioned into one generative model.
Constructing time-domain representations of unmodelled glitches is challenging, but made possible using algorithms such as \textit{BayesWave} to isolate them from the detector background.
Once accurate glitch representations of other LIGO glitch types are constructed, cDVGAN is scheduled for further development in a next-generation glitch generator.
Covering the entire LIGO glitch space with cDVGAN will result in a model more representative of LIGO glitches and a useful tool for downstream analysis.

\section{Acknowledgements}
This research was conducted within the ET Technologies project (PROJ-03612) which is partly funded by EFRO, the Province of Limburg and The Dutch Ministry of Economic Affairs and Climate Policy within the REACT-EU Programme of OP Zuid.
The authors are grateful for contributions by members of the ET Technologies research team, in particular; 
Stefano Schmidt, Andrew Miller and Sarah Caudill.
This material is based upon work supported by NSF's LIGO Laboratory which is a major facility fully funded by the National Science Foundation.
The authors are grateful for computational resources provided by the LIGO Laboratory and supported by the National Science Foundation Grants No. PHY-0757058 and No. PHY-0823459.

\newpage

\onecolumngrid

\newpage
\section{Appendix}

\vspace{-3mm}

\subsection{\label{sec:cDVGAN_Architecture}cDVGAN Architecture}

\begin{table}[H]
\centering
\begin{tabular}{llllll} 
    \toprule
    {} & {} & {Discriminator} & {} & {} & {(3.5M param.)} \\ \hline
    {Operation} & {Output shape} & {Kernel size} & {Stride} & {Dropout} & {Activation} \\ \hline
    {Input} & (1024) & {-} & {-}  & {0}  & {-}\\ 
    {Reshape} & (64,16) & {-} & {-}  & {0}  & {-}\\ 
    {Convolutional} & (64,128) & {14} & {2}  & {0.5}  & {Leaky ReLU}\\ 
    {Convolutional} & (32,128) & {14} & {2}  & {0.5}  & {Leaky ReLU}\\ 
    {Convolutional} & (16, 256) & {14} & {2}  & {0.5}  & {Leaky ReLU}\\ 
    {Convolutional} & (8, 256) & {14} & {2}  & {0.5}  & {Leaky ReLU}\\ 
    {Convolutional} & (4, 512) & {14} & {2}  & {0.5}  & {Leaky ReLU}\\ 
    {Global Avg. Pooling} & (512) & {-} & {-}  & {0.5}  & {-}\\ 
    {Avg. Pooling Dense} & (128) & {-} & {-}  & {0.2}  & {Leaky ReLU}\\
    {Dense} & (1) & {-} & {-}  & {0}  & {Linear}\\
    {Class Input} & (3) & {-} & {-}  & {-}  & {-}\\
    {Class Dense} & (128) & {-} & {-}  & {0}  & {Linear}\\
    {Scalar Product} & (1) & {-} & {-}  & {-}  & {-}\\     
    {Dense + Scalar Product} & (1) & {-} & {-}  & {-}  & {-}\\ \bottomrule
    
    {} & {} & {DV Discriminator} & {} & {} & {(1.1M param.)} \\ \hline
    {Operation} & {Output shape} & {Kernel size} & {Stride} & {Dropout} & {Activation} \\ \hline
    {Input} & (1023) & {-} & {-}  & {0}  & {-}\\ 
    {Dense} & (512) & {-} & {-}  & {0}  & {Leaky ReLU}\\ 
    {Reshape} & (32,16) & {-} & {-}  & {0}  & {-}\\ 
    {Convolutional} & (32, 64) & {5} & {2}  & {0.5}  & {Leaky ReLU}\\ 
    {Convolutional} & (16,128) & {5} & {2}  & {0.5}  & {Leaky ReLU}\\ 
    {Convolutional} & (8,256) & {5} & {2}  & {0.5}  & {Leaky ReLU}\\ 
    {Convolutional} & (4,256) & {5} & {2}  & {0.5}  & {Leaky ReLU}\\ 
    {Global Avg. Pooling} & (256) & {-} & {-}  & {0.5}  & {-}\\
    {Avg. Pooling Dense} & (128) & {-} & {-}  & {0.2}  & {Leaky ReLU}\\
    {Dense} & (1) & {-} & {-}  & {0}  & {Linear}\\
    {Class Input} & (3) & {-} & {-}  & {-}  & {-}\\
    {Class Dense} & (128) & {-} & {-}  & {0}  & {Linear}\\
    {Scalar Product} & (1) & {-} & {-}  & {-}  & {-}\\     
    {Dense + Scalar Product} & (1) & {-} & {-}  & {-}  & {-}\\ \bottomrule
    
    {} & {} & {Generator} & {} & {} & {3.5M param.} \\ \hline
    {Operation} & {Output shape} & {Kernel size} & {Stride} & {BN} & {Activation} \\ \hline
    {Latent input} & (100) & {-} & {-}  & {\xmark}  & {-}\\
    {Class Input} & (3) & {-} & {-}  & {\xmark}  & {-}\\
    {Class Dense} & (32) & {-} & {-}  & {\xmark}  & {-}\\
    {Concatenate} & (132) & {-} & {-}  & {\xmark}  & {-}\\
    {Dense} & (1024) & {-} & {-}  & {\xmark}  & {ReLU}\\ 
    {Reshape} & (32,32)  & {-} & {-}  & {\xmark}  & {-}\\ 
    {Transposed conv.} & (64,512) & {18} & {2}  & {\cmark}  & {ReLU}\\ 
    {Transposed conv.} & (128,256) & {18} & {2}  & {\cmark}  & {ReLU}\\ 
    {Transposed conv.} & (256,128) & {18} & {2}  & {\cmark}  & {ReLU}\\ 
    {Transposed conv.} & (512, 64) & {18} & {2}  & {\cmark}  & {ReLU}\\ 
    {Transposed conv.} & (1024,1)& {18} & {2}  & {\xmark}  & {Linear}\\ 
    {Flatten} & (1024) & {-} & {-}  & {\xmark}  & {-} \\ \hline
    {Optimizer} & {RMSprop($\alpha = 0.0001$)} & {} & {} & {} & {} \\
    {Batch size} & {512} & {} & {} & {} & {} \\
    {Epochs} & {500} & {} & {} & {} & {} \\
    {Loss} & {Wasserstein} & {} & {} & {} & {} \\ \bottomrule
\end{tabular}
\captionsetup{width = 0.85 \textwidth}
\caption{The architecture and hyperparameters describing cDVGAN, which consists of a base discriminator, a derivative (DV) discriminator and a generator convolutional network. The additional discriminator of cDVGAN2 follows the same architecture as the DV disciminator but with an input shape of 1022. The number of parameters (param.) in each model are shown in brackets beside the name of each model component.}
\label{tab:model_architecture}
\end{table}

\subsection{\label{sec:CNN_Architecture}CNN Architecture}

\begin{table*}[!htb]
\centering
\begin{tabular}{llllll} 
    \toprule
    {Operation} & {Output shape} & {Kernel size} & {Stride} & {Dropout} & {Activation} \\ \hline
    {Input} & (1024) & {-} & {-}  & {0}  & {-}\\ 
    {Reshape} & (1024,1) & {-} & {-}  & {0}  & {-}\\ 
    {Convolutional} & (512,256) & {5} & {2}  & {0.5}  & {Leaky ReLU}\\ 
    {Convolutional} & (256,128) & {5} & {2}  & {0.5}  & {Leaky ReLU}\\
    {Convolutional} & (128,64) & {5} & {2}  & {0.5}  & {Leaky ReLU}\\
    {Convolutional} & (64,32) & {5} & {2}  & {0.5}  & {Leaky ReLU}\\
    {Flatten} & (2048) & {-} & {-}  & {0.5}  & {-}\\ 
    {Dense} & (512) & {-} & {-}  & {0}  & {Leaky ReLU}\\
    {Dense} & (1) & {-} & {-}  & {0}  & {Sigmoid}\\ \bottomrule
    
    {Optimizer} & {Adam($\alpha = 0.001$)} & {} & {} & {} & {} \\
    {Batch size} & {64} & {} & {} & {} & {} \\
    {Epochs} & {20} & {} & {} & {} & {} \\
    {Loss} & {Binary Crossentropy} & {} & {} & {} & {} \\ \bottomrule
\end{tabular}
\captionsetup{width = 0.85 \textwidth}
\caption{The architecture of the CNN (1.2M parameters) used during experiments.}
\label{tab:CNN_model_architecture}
\end{table*}

\subsection{\label{sec:Standard_class_hybrid}Vertex, Simplex and Uniform Datasets}

\begin{figure}[!htb]
\centering
\captionsetup{width = 0.8 \textwidth}
  \includegraphics[width=0.7\textwidth]{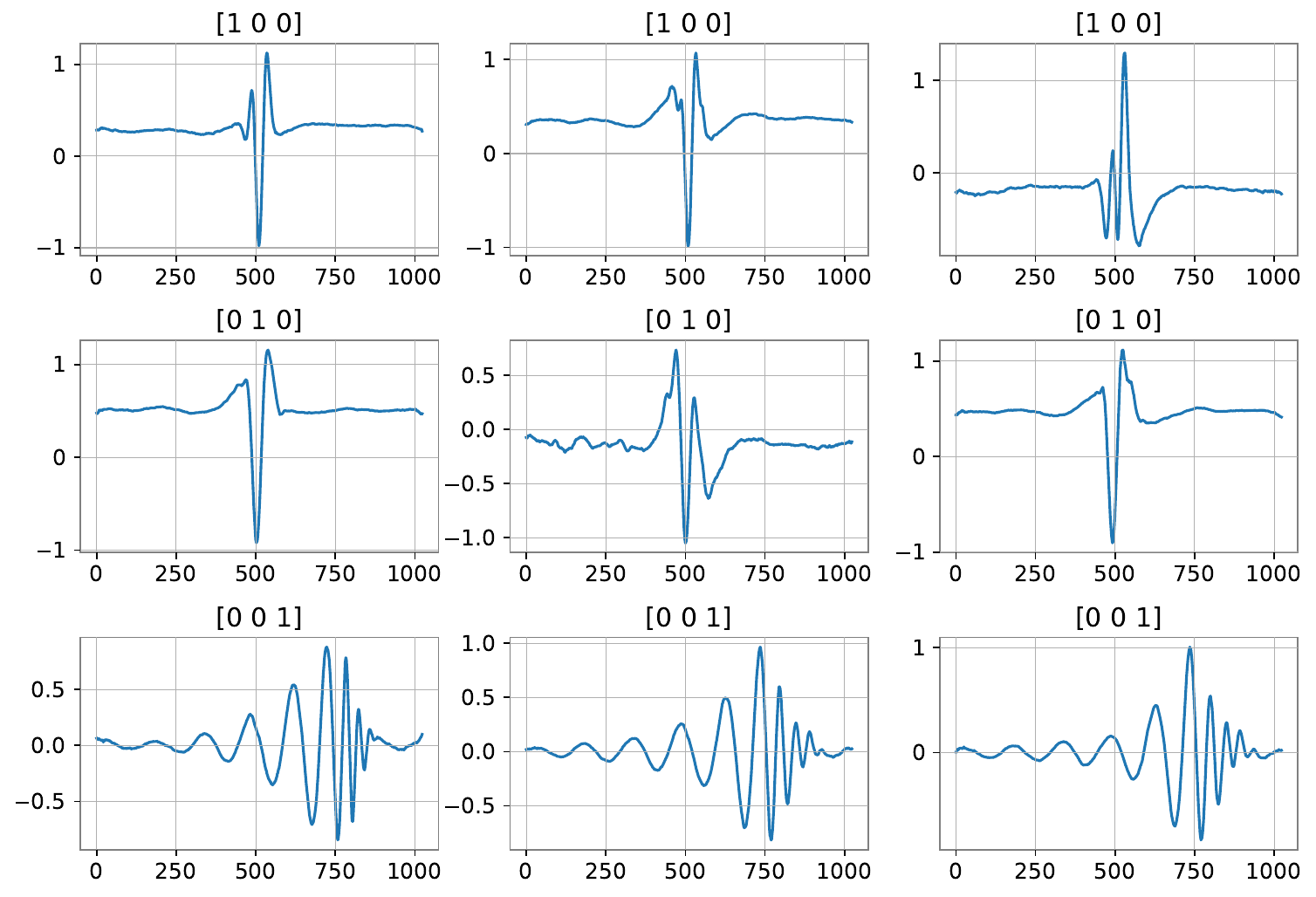}
\caption{Standard 3-class vertex generations from cDVGAN.}
\label{fig:vertex_generations}
\end{figure}

\begin{figure}[!htb]
\centering
\captionsetup{width = 0.8 \textwidth}
  \includegraphics[width=0.7\textwidth]{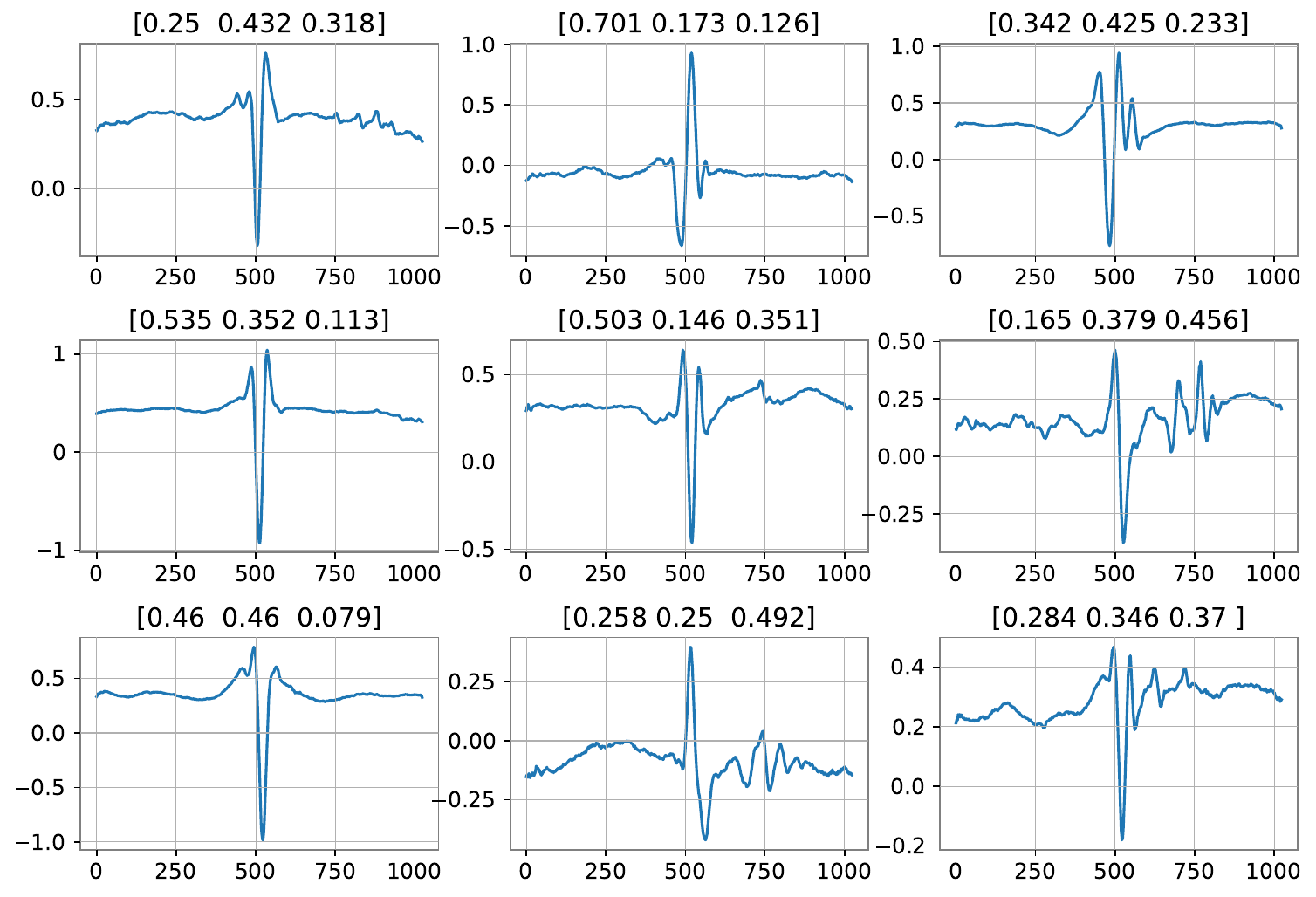}
\caption{Simplex generations from cDVGAN.}
\label{fig:simplex_generations}
\end{figure}

\begin{figure}[!htb]
\centering
\captionsetup{width = 0.8 \textwidth}
  \includegraphics[width=0.7\textwidth]{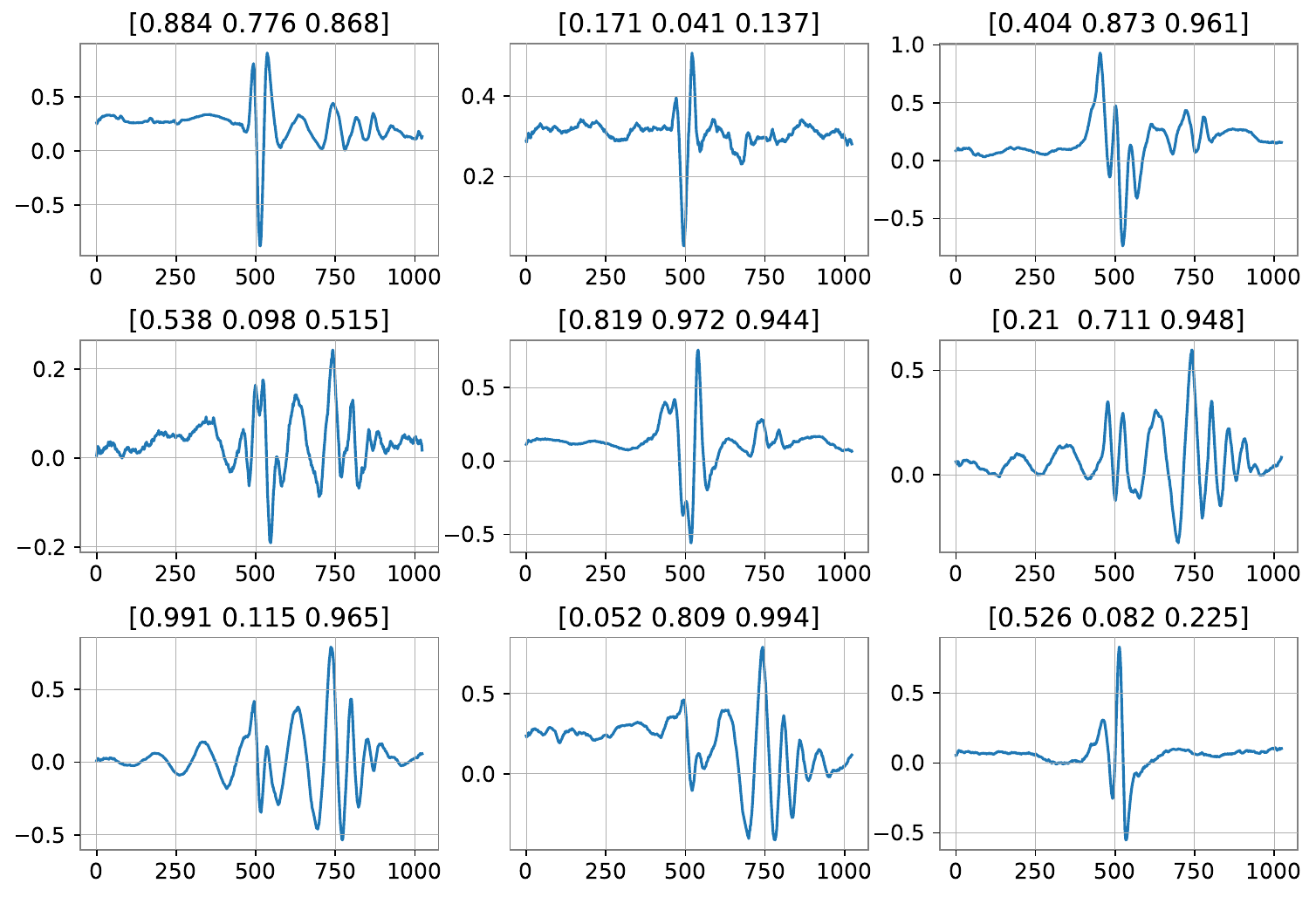}
\caption{Uniform generations from cDVGAN.}
\label{fig:uniform_generations}
\end{figure}


\clearpage

\subsection{\label{sec:Class_Interpolation}Class Interpolation}

\begin{figure}[!htb]
\centering
\captionsetup{width = 0.8 \textwidth}
  \includegraphics[width=\textwidth]{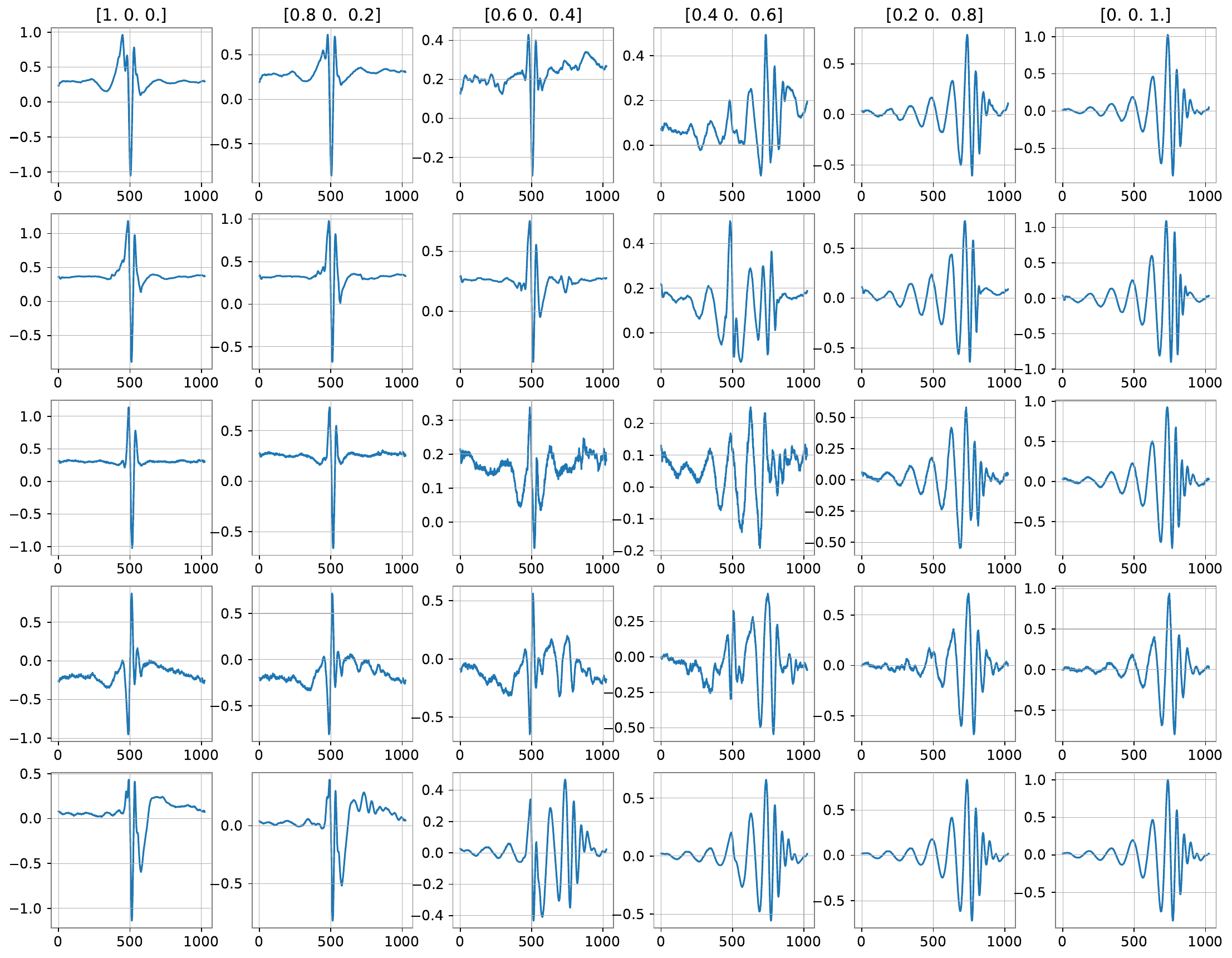}
\caption{Interpolation between blip and BBH classes for cDVGAN (1st row), cDVGAN2 (2nd row), cWGAN (3rd row), McGANn (4th row) and McDVGANn (5th row). The class input is shown at the top of each column, while the latent input of the generator is kept constant.}
\label{fig:interpolation_simplex}
\end{figure}

\newpage\subsection{\label{sec:Gspy_analysis}Gravity Spy Analysis of GAN Training Data}

As we have discussed in section \ref{sec:CNN_classifier}, we have scaled some examples of blip and tomte glitches $g(t)$, adding them to the noise $n(t)$ from the third observing (O3) run in the GPS time range $[1262540000, 1262540040]$. Note that the glitches have been extracted from their original noise following the procedure previously presented in section \ref{sec:Data_preprocessing}.  Afterwards, the time-series $s(t) = g(t) + n(t)$ was classified with Gravity Spy, providing a class label and a class confidence $c_{GS}$. 

\begin{figure}[h]
\centering
\captionsetup{width = 1 \textwidth}
\begin{subfigure}[t]{0.33\textwidth}
  \centering
  \includegraphics[width = \textwidth]{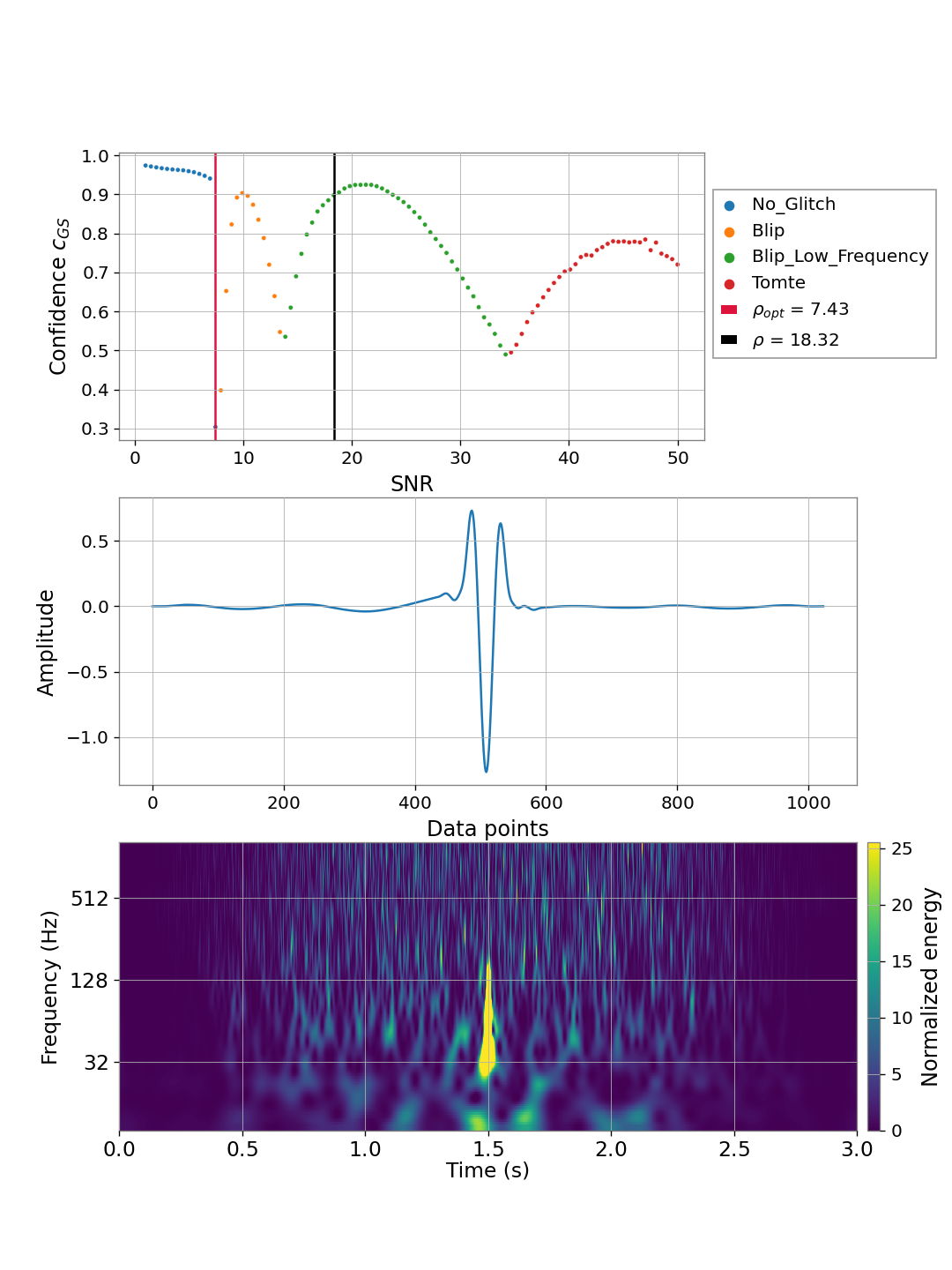}
  \caption{}
  \label{fig:glitch_0}
\end{subfigure}%
\begin{subfigure}[t]{0.33\textwidth}
  \centering
  \includegraphics[width = \textwidth]{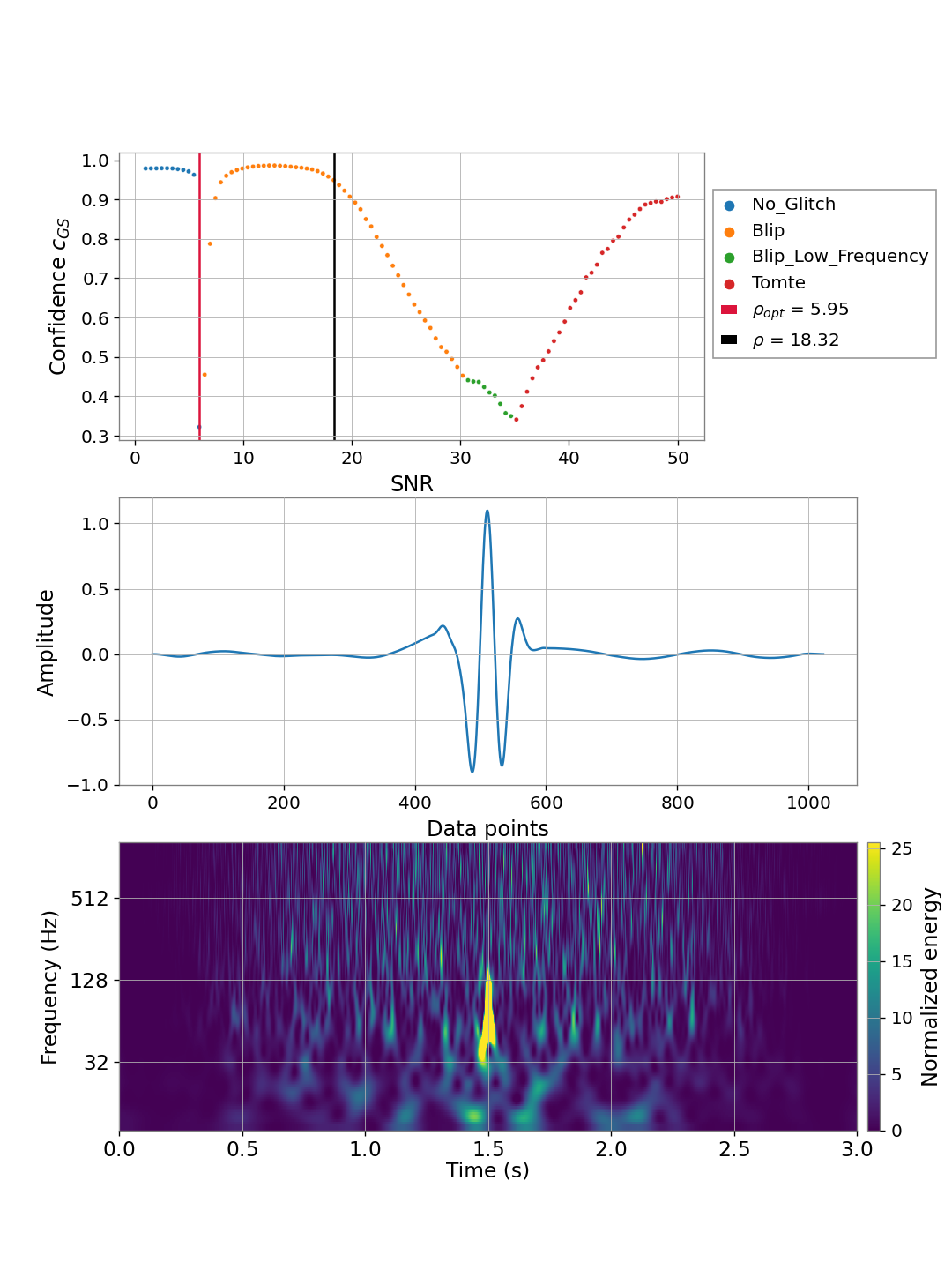}
  \caption{}
  \label{fig:glitch_1}
\end{subfigure}%
\begin{subfigure}[t]{0.33\textwidth}
  \centering
  \includegraphics[width = \textwidth]{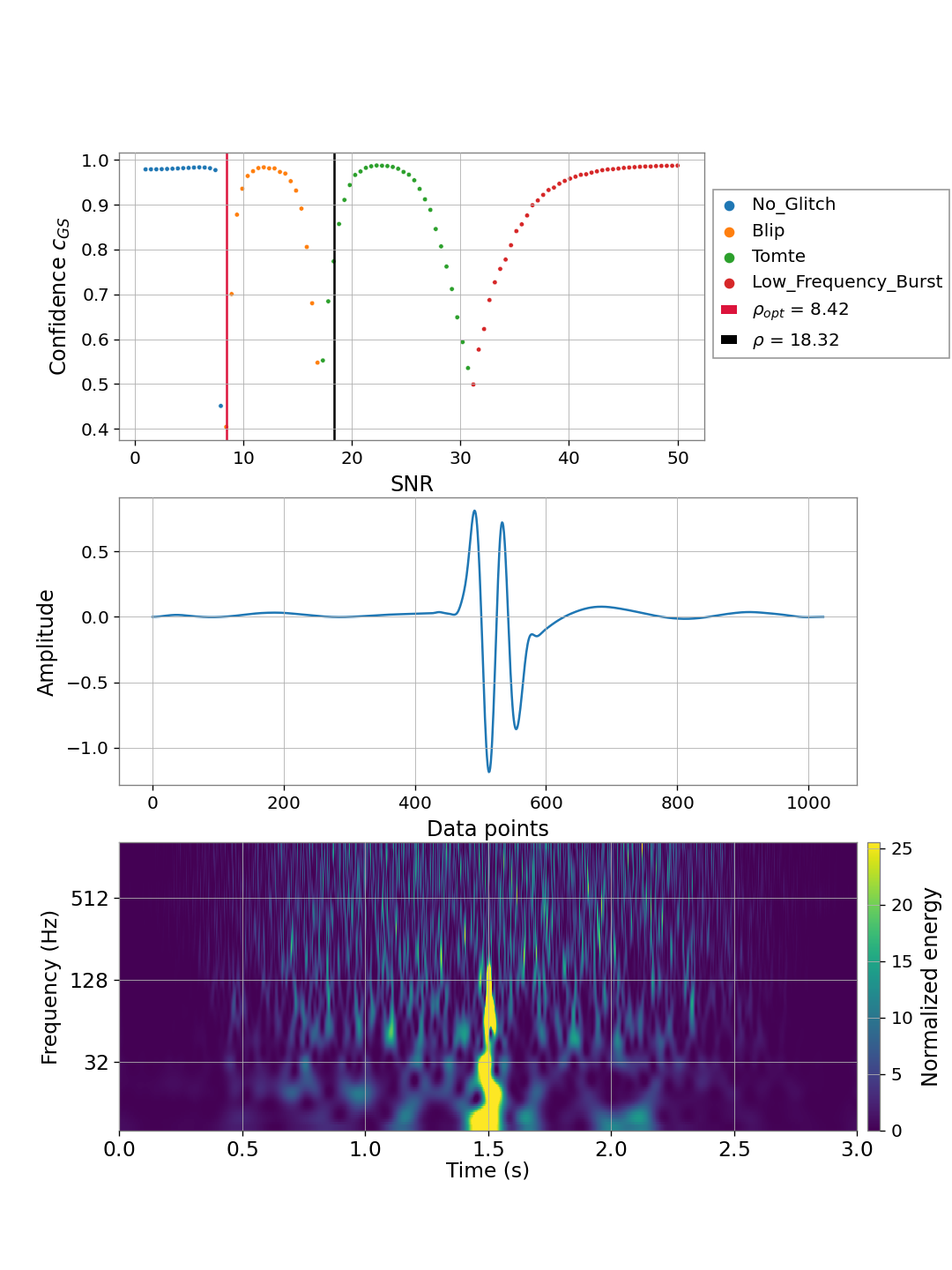}
  \caption{}
  \label{fig:glitch_2}
\end{subfigure}%
\caption{\textit{Top panel:} Three examples of blip glitches classified by \textit{Gravity Spy}, with the classification confidences as a function of SNR. \textit{Middle panel:} The time-series representation of the classified glitch. \textit{Bottom panel:} The corresponding spectrogram representation of the classified glitch after injecting into whitened detector noise.\label{fig:gspy_blip}} 
\end{figure}

In Fig. \ref{fig:gspy_blip} we present the results of classifying three different denoised blips re-injected in O3 noise. In the top panel, we plot the the confidence $c_{GS}$ as a function of the optimal SNR $\rho_{opt}$, defined in Eq. \ref{eq:snr_scaling}. In the middle panel, we show the time-series of the blip as outputted by the pre-processing, and in the bottom panel, we show the spectrogram or time-frequency representation of the glitch embedded in O3 real noise at $\rho_{opt} = 18.32$.

 We can observe in Fig. \ref{fig:gspy_blip} that \textit{Gravity Spy} classifies the time-series as No$\_$Glitch for SNR $\lesssim 7$, which is expected as this algorithm only learns glitches with an SNR $\leq 7.5$. While these examples are labelled as blips at SNR $\sim 10$, as the SNR increases they get misclassified as Blip$\_$Low$\_$Frequency or Low$\_$Frequency$\_$Burst,  which can be explained by the use of band-pass filtering that attenuates the high-frequency contribution of blips.
 
\begin{figure}[h]
\centering
\captionsetup{width = 0.9\textwidth}
\begin{subfigure}[t]{0.33\textwidth}
  \centering
  \includegraphics[width = \textwidth]{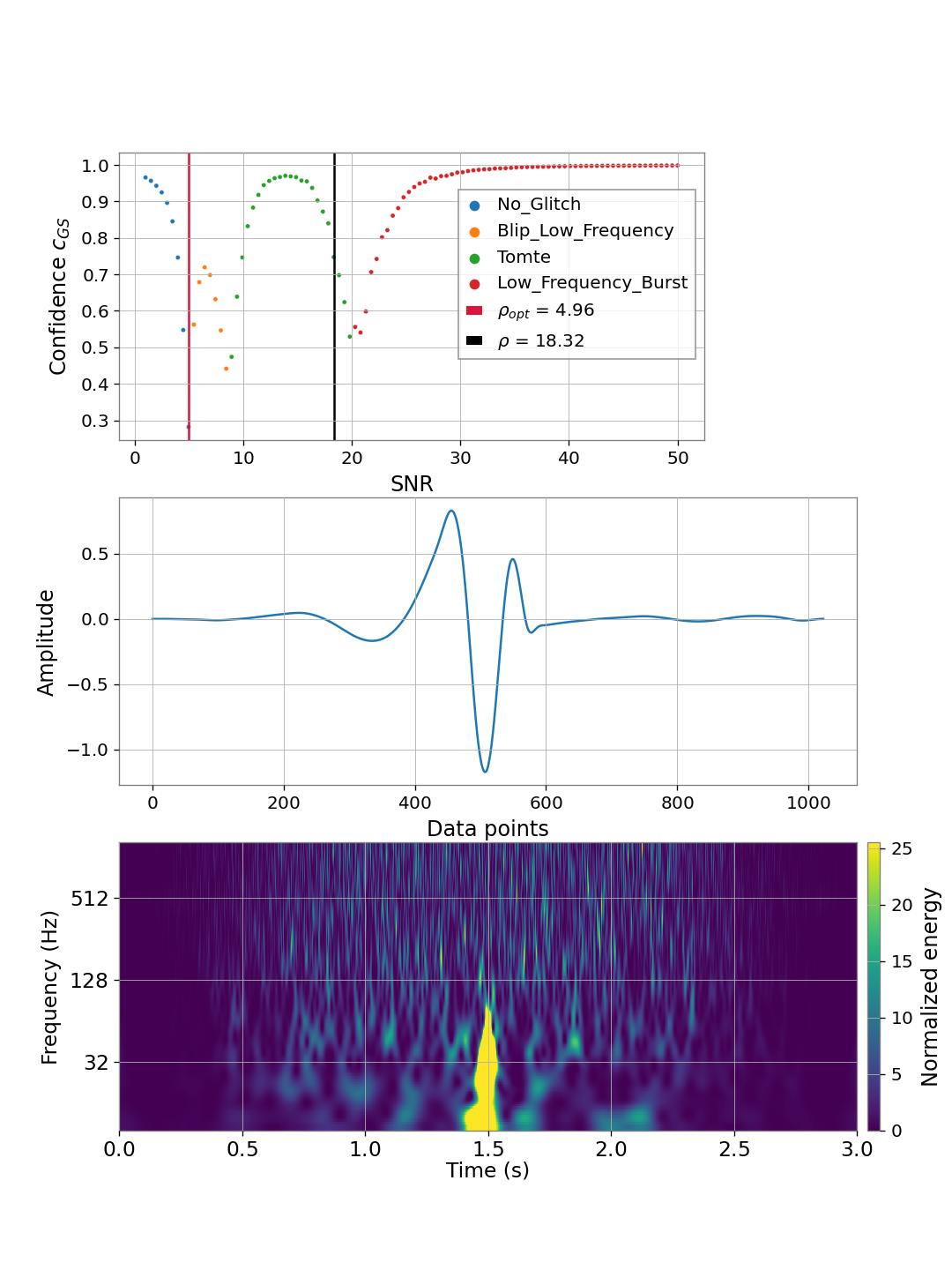}
  \caption{}
  \label{fig:glitch_20}
\end{subfigure}%
\begin{subfigure}[t]{0.33\textwidth}
  \centering
  \includegraphics[width = \textwidth]{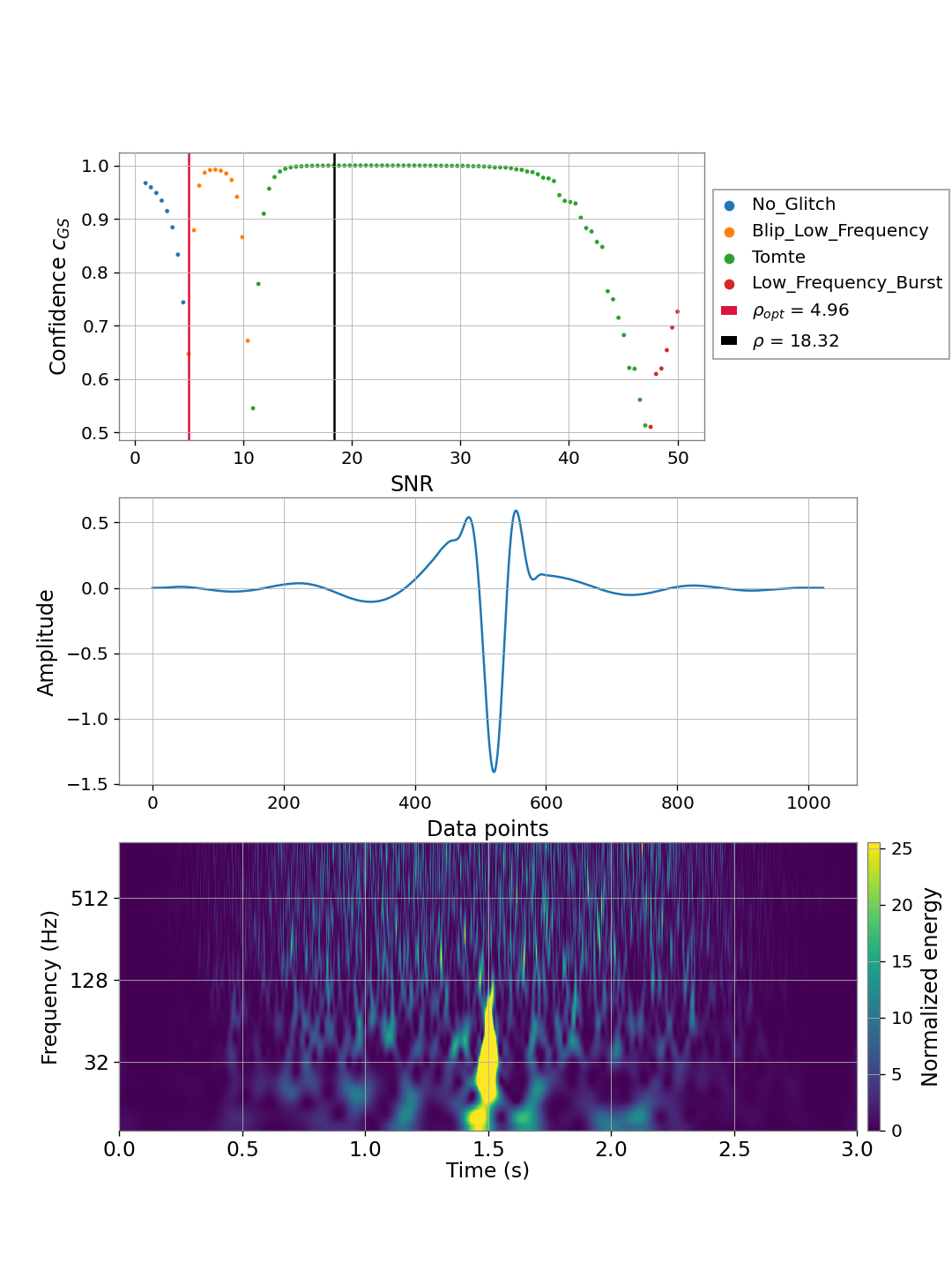}
  \caption{}
  \label{fig:glitch_21}
\end{subfigure}%
\begin{subfigure}[t]{0.33\textwidth}
  \centering
  \includegraphics[width = \textwidth]{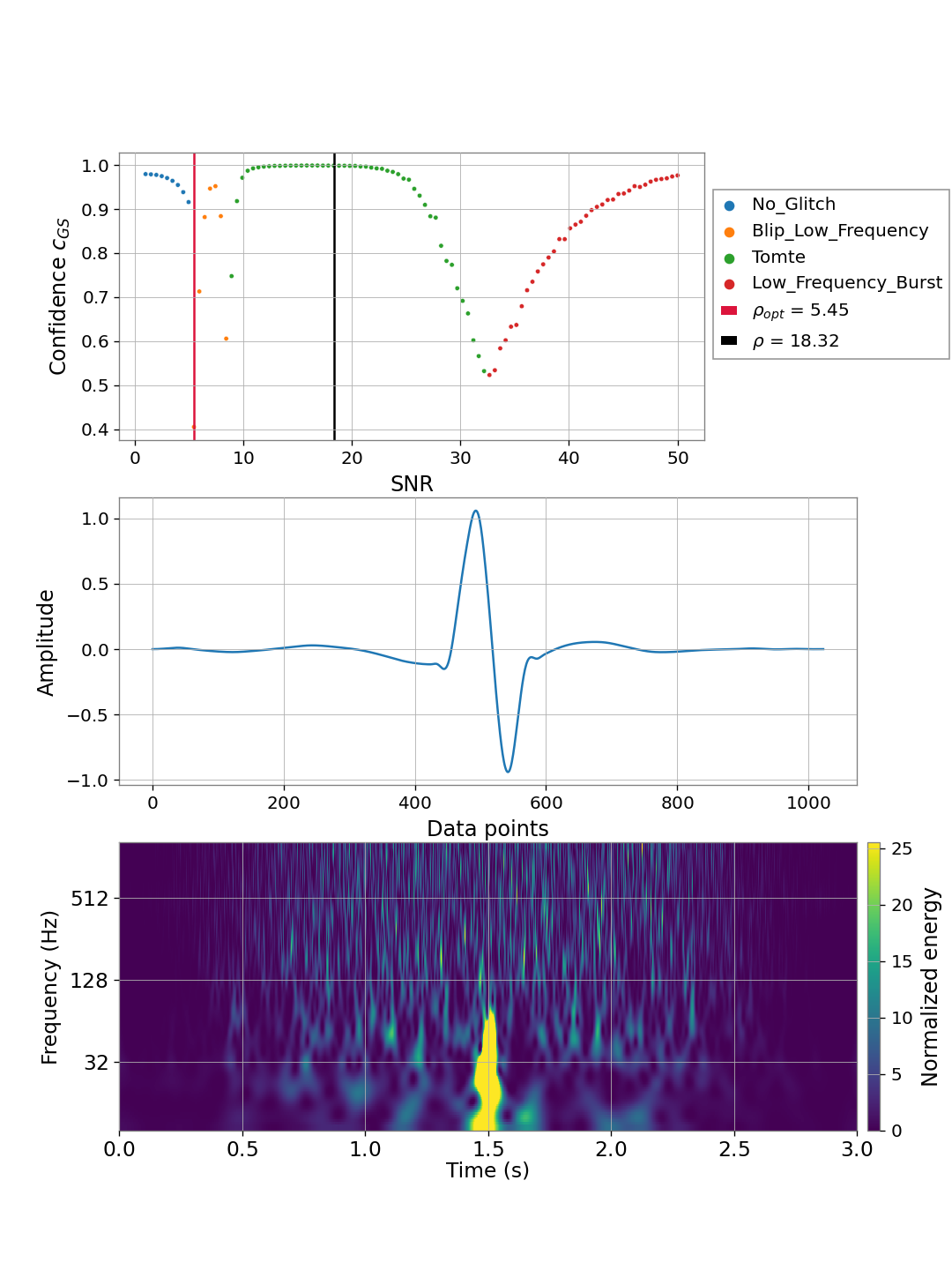}
  \caption{}
  \label{fig:glitch_23}
\end{subfigure}%
\caption{\textit{Top panel:} Three examples of tomte glitches classified by \textit{Gravity Spy}, with the classification confidences as a function of SNR. \textit{Middle panel:} The time-series representation of the classified glitch, \textit{Bottom panel:} The corresponding spectrogram representation of the classified glitch after injecting into whitened detector noise.}
\label{fig:gspy_tomte}
\end{figure}

Similarly, in Fig. \ref{fig:gspy_tomte} we can also observe that the time-series gets classified as No$\_$Glitch, but this time at an SNR $\lesssim 5$. Then, for $5 \lesssim \text{ SNR }< 10$ it gets briefly classified as Blip$\_$Low$\_$Frequency, to then be classified as Tomte for most of the SNR range, meaning that the features of tomtes are better preserved than the blip class after data pre-processing.

\newpage\subsection{\label{sec:McDVGANn_1}McDVGANn Additional Analysis}

\begin{figure}[!htb]
\centering
\captionsetup{width = 0.8 \textwidth}
  \includegraphics[width=0.5\textwidth]{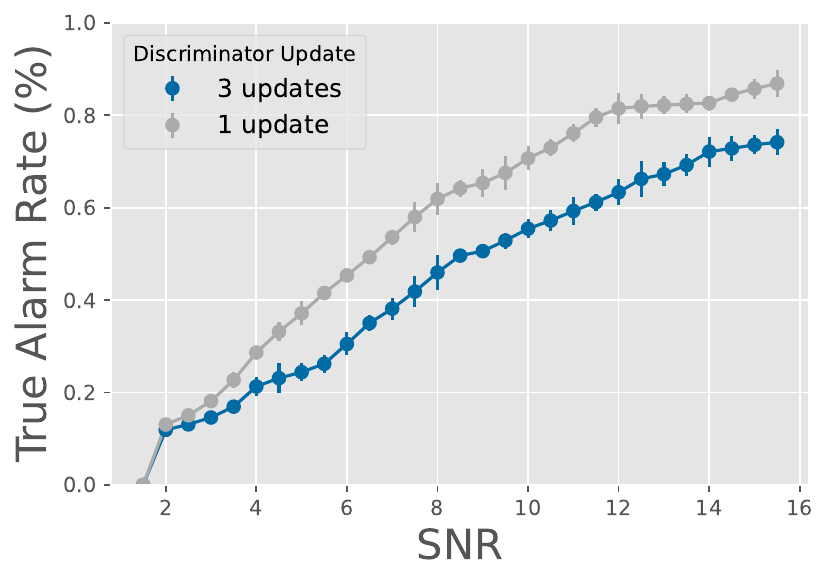}
\caption{A comparison of McDVGANn CNN performance trained with a discriminator update schedule of 1 for each generator update and a discriminator update schedule of 3 times for each generator update, which is used for the original McGANn model. These results suggest that McDVGANn overfits the data with the original McGANn update schedule and that using a discriminator update schedule of 1 is superior.}
\label{fig:mcdvgann_snr_vs_tap}
\end{figure}

\newpage
\twocolumngrid

\bibliography{apssamp}

\end{document}